\documentclass[iop,appendixfloats]{emulateapj}

\usepackage[usenames,dvipsnames]{color}
\usepackage{color}
\usepackage{amsmath}
\usepackage{float}
\usepackage{ctable}
\usepackage{graphicx}
\usepackage{refcount}
\usepackage{derivative}
\usepackage{lineno,hyperref}
\usepackage[normalem]{ulem}
\newcommand{\figurepath}{.}

\newcommand{\cs}{c_{\rm s}}

\newbox\grsign \setbox\grsign=\hbox{$>$}
\newdimen\grdimen \grdimen=\ht\grsign
\newbox\laxbox \newbox\gaxbox
\setbox\gaxbox=\hbox{\raise.5ex\hbox{$>$}\llap
     {\lower.5ex\hbox{$\sim$}}}\ht1=\grdimen\dp1=0pt
\setbox\laxbox=\hbox{\raise.5ex\hbox{$<$}\llap
     {\lower.5ex\hbox{$\sim$}}}\ht2=\grdimen\dp2=0pt

\shorttitle{Dipole vs. disk Fields in disk-jet}
\shortauthors{}
\begin{document}
\title{Interaction of Stellar Dipole and Disk Magnetic Fields in Accretion-Ejection Systems}
\author{Somayeh Sheikhnezami
\altaffilmark{1}
\&
        Sarah Aslani
\altaffilmark{1}
}
\altaffiltext{1}{Department of Physics, Institute for Advanced Studies in Basic Sciences (IASBS), Zanjan 45137-66731, Iran}
\email{snezami@iasbs.ac.ir, snezami@ipm.ir}

     \date{\today}
\begin{abstract}

We study jet formation from a magnetically diffusive, sub-Keplerian accretion disk threaded by different magnetic field configurations.
We present two dimensional, axisymmetric MHD simulations in spherical coordinates employing the PLUTO code.
Our results show that the reference simulation, in which the disk is threaded by an initial large scale poloidal magnetic field, reaches a quasi steady state and produces a well collimated jet.
Both kinetic and magnetic torques are significant and evolve smoothly over time.
The accretion process is established smoothly, with about 26\% of the accreting material being delivered into the outflow.
Alternatively, applying a purely stellar dipolar magnetic field produces weak and unstable outflows, characterized by inefficient angular momentum transport and strongly suppressed accretion.
Additionally, in simulations where the stellar dipolar magnetic field is present alongside the disk's initial poloidal magnetic field, the disk magnetic field predominantly stabilizes the system.
It also facilitates angular momentum transport from the disk and sustains a significant outflow mass flux.
Furthermore, we performed additional simulations that explicitly include an inner gap associated with the disk truncation region.
These simulations were designed as a robustness test of the treatment of the innermost disk region and confirm that our main conclusions are not sensitive to the presence of this gap.
Overall, our findings demonstrate that although the magnitude of stellar dipole plays an important role in the evolution of the accretion-ejection structure, the large scale poloidal magnetic field threading the disk is essential for launching and sustaining a stable and well collimated magnetized jet.
\end{abstract}

\keywords{ Protostellar disk, Magnetic fields, Magnetohydrodynamics (MHD), MHD simulation, ISM: jets and outflows }

\section{Introduction}
One of the important phases of star formation is the launching of stellar jets from the circumstellar disk.
These are high velocity, collimated outflows of material ejected from the underlying disk and play an essential role in regulating star formation by removing angular momentum and facilitating mass accretion onto the protostar.
Among the various jet formation scenarios, the magneto centrifugal acceleration mechanism introduced by \cite{1982MNRAS.199..883B} is one of the most widely accepted.
The overall idea is that energy and angular momentum are extracted from the disk through an efficient magnetic torque, which arises from a large scale magnetic field threading the disk.

Based on this concept, two classes of models have been developed, the "MHD disk wind" \citep{1983ApJ...274..677P} and the ``X-wind'' \citep{1994ApJ...429..781S}.
Basically, these models differ in the location of the jet launching region and in their assumptions regarding the magnetic field configuration and boundary conditions at the star-disk interface~\cite{2008ApJ...672..489C}.

In the MHD disk-wind framework, the disk is threaded by an open, large scale magnetic field, which exerts a torque on the disk and an outflow over an extended radial range is formed.
The physical processes that drive the outflow, the conditions necessary for steady jet launching, and the properties of magnetized accretion-ejection structures have been extensively studied in previous theoretical works~\citep{1993A&A...276..625F,1993A&A...276..637F,1995A&A...295..807F}.
Subsequently, \cite{1997A&A...319..340F} investigated the continuous solutions covering all dynamical processes, in stationary, weakly dissipative magnetized accretion-ejection structures of bipolar topology extending from the accretion disk to super Alfvénic jets.

Numerous studies, have also focused on jet formation through numerical simulations based on the MHD disk-wind framework.
Pioneering work, in this area has been conducted by \citet{1985PASJ...37...31S,1995ApJ...439L..39U,1997ApJ...482..712O}.
In these works, the disk evolution is not considered in the numerical treatment and the jet is formed from a slow disk-wind injected from the disk surface, where the disk is effectively treated as a boundary condition.
This approach is numerically less expensive and enables a wide range of parameter studies.
This approach has subsequently been developed by incorporating additional physical processes and examining their impact on jet acceleration and collimation.
For instance, self-collimated MHD jets for low mass stars have been studied~\citep{1997ApJ...482..712O,1999ApJ...526..631K}. Additionally, other studies have investigated the role of magnetic diffusivity~\citep{2002A&A...395.1045F} and radiative forces~\citep{2011ApJ...742...56V}.

The MHD disk-wind model of jet launching from the disk boundary has also been applied to relativistic outflows~\citep{2010ApJ...709.1100P}.
Furthermore, it has been extended to explore non-axisymmetric three-dimensional (3D) effects on jet formation~\citep{2003ApJ...582..292O,2006ApJ...653L..33A} and to consider polarized synchrotron radiation transfer in relativistic jets using 3D MHD simulations~\citep{2013MNRAS.429.2482P}.
More recently, magnetized jet formation from the disk boundary has been further studied by considering 3D and tidal effects induced by a companion star~\citep{2024ApJ...966...82S}.

Further studies extended this framework by explicitly including disk evolution within the accretion-ejection system.
The accretion process in magnetized disks was first studied through numerical simulations by~\cite{1994ApJ...433..746S}. Further pioneering work was presented by~\cite{1998ApJ...508..186K}, who were the first to perform simulations of jet launching from a diffusive MHD disk.
This approach was later extended to considerably longer timescales and larger spatial scales~\citep{2002ApJ...581..988C,2004ApJ...601...90C,2007A&A...469..811Z,2012ApJ...757...65S,2013ApJ...774...12F}.
This approach enabled the simulations to reach steady states and provided better estimates of the corresponding mass fluxes.

Following this approach, further physical effects were investigated, including the influence of disk magnetization \citep{2009MNRAS.400..820T}, launching from viscous disks \citep{2010A&A...512A..82M}, and thermal effects \citep{2013MNRAS.428.3151T}.
In addition, the launching of outflows from magnetic fields self-generated by a mean-field disk dynamo has been examined in several studies~\citet{2003A&A...398..825V,2004A&A...420...17V,2014ApJ...793...31S,2014ApJ...796...29S,2018ApJ...855..130F,2022ApJ...935...22M}.
Some works have also extended this approach to the launching of relativistic outflows in the context of GRMHD mean-field disk dynamos~\citep{2021ApJ...911...85V}.

Moreover, the problem of jet launching and the associated accretion-ejection process in 3D simulations has been studied recently~\citep{2015ApJ...814..113S,2018ApJ...861...11S,2022ApJ...925..161S}.
These studies include disk evolution and examine non-axisymmetric 3D effects, such as tidal interactions induced by a companion, providing a more complete view of the physical processes governing the accretion-ejection structure.
More recently, \citet{2025ApJ...986...51S} presented an even more advanced study, performing 3D spherical simulations of a protostellar disk the launched jet in a binary system, coupled with post-processed 3D radiative transfer modeling using RADMC-3D~\citep{2012ascl.soft02015D}.
This approach enables a direct connection between the simulated disk-jet structures and their observable signatures in dust continuum emission, as detectable with JWST or ALMA.

In addition to the large scale magnetic field threading the disk, commonly employed in most of the aforementioned studies, the central protostar also generates a dipolar magnetic field that can significantly influence both the launching process and the resulting outflow structure.
The earliest simulations exploring the impact of a dipolar magnetic field on the surrounding accretion disk were performed by~\citet{1984PASJ...36..105U,1985PASJ...37..515U}.
The model describing the interaction between a dipolar magnetic field and the surrounding disk, along with the resulting outflows and accretion funnels, was extensively studied by~\cite{1994ApJ...429..781S} and is known as the X-wind model.
In this framework, the wind is launched from the X-region at the inner edge of the disk, where the stellar magnetosphere dominates and material is transferred along magnetic field lines.
Following the X-wind model, further studies investigated the details of the interaction between the stellar dipolar magnetic field and the circumstellar disk.
\cite{1997ApJ...489..890M} performed time dependent simulations of dipolar magnetospheres, initialized with either a pure dipolar magnetic field or a dipole field superposed with a vertical constant magnetic field. However, the timescale of these simulations, which included disk evolution, was short, only a few inner disk rotations.
Furthermore, \citet{1999ApJ...524..142G,2002ApJ...574..232M} presented simulations of dipolar star-disk magnetospheres for up to 150 inner disk rotations. These simulations exhibit magnetic reconnection and flaring behavior in the vicinity of the inner disk radius, as well as  the formation of a highly collimated, narrow jet along the rotation axis.
In comparison, \citet{1999A&A...349L..61F,2000A&A...363..208F} did not explicitly model the disk structure in their simulations, which allowed them to follow the evolution of the stellar magnetosphere over more than 2000 inner disk rotation periods. Their results demonstrated, for the first time, that the axial jet feature reported in some earlier studies is intermittent and does not persist over long time intervals.

Additionally, numerous studies have studied how a dipolar magnetic configuration affects accretion processes, drives outflow formation, and in some runs, includes the evolution of the surrounding disk~\citep{2002ApJ...578..420R,2003Ap&SS.287...65M,2003ApJ...595.1009R,2008ApJ...678.1109M,2009A&A...508.1117Z,2013A&A...550A..99Z,2021ApJ...906....4I,2023A&A...678A..57C,2023MNRAS.523.4708M}.
For example, \citet{2002ApJ...578..420R,2003ApJ...595.1009R} employed axisymmetric and 3D ideal MHD simulations of purely dipolar magnetic fields to show how a star gains angular momentum through magnetic stresses, leading to stellar spin-up.
Moreover, \citet{2009A&A...508.1117Z} presented simulations of viscous, resistive accretion disk interacting with a stellar dipolar magnetosphere, finding that star-disk interaction removes only ~10\% of the angular momentum, while stellar winds contribute ~20\% due to low mass-loss rates, insufficient to counteract accretion-driven spin-up.
Consistent with this, \citet{2011ApJ...727L..22Z} showed that accretion powered stellar winds alone cannot explain the slow rotation of classical T Tauri stars, reinforcing that stellar winds are unlikely to be the sole spin-down mechanism.
\citet{2023A&A...678A..57C} showed that a stellar dipolar field is suppressed inside thin disks but influences angular momentum transport via field inflation outside the disk. \citet{2024MNRAS.528.2883Z} further showed that magnetic interchange instabilities generate complex accretion structures and weak outflows in 3D magnetospheric accretion onto a non-rotating star.

The interaction between stellar and disk magnetic fields was further investigated in several studies~\citep{2009ApJ...692..346F,2009A&A...502..217M}.
Adopting a two-component magnetic field configuration, these authors explored the coupling among the stellar magnetosphere, stellar wind, and magnetized disk outflow.
\citet{2009ApJ...692..346F} discussed the importance of field alignment for outflow structure and showed that, while differential rotation drives launching, a strong disk wind is required for effective jet collimation.
The disk was treated as a fixed boundary to enable high resolution modeling of the outflow over a large computational domain.

Overall, despite the conceptual appeal of the X-wind scenario, the model remains debated.
No global simulations have yet provided clear numerical confirmation of a classical X-wind in operation.
In practice, such models are often interpreted in the literature as disk winds launched from a limited radial range near the inner disk rather than as a distinct X-wind solution.
Simulations that include only a purely dipolar stellar magnetosphere fail to produce a persistent, well collimated jet, indicating that an additional disk-wind component is required.
This suggests a ``two-component outflow'' scenario for jet formation.

Motivated by the above discussion, we investigate how stellar and disk magnetic field configurations influence jet launching from a diffusive accretion disk.
Specifically, we examine three magnetic field configurations:
(i) an initially large scale poloidal magnetic field threading the disk,
(ii) a purely stellar dipolar field, and
(iii) the superposition of both fields.

It should be emphasized that these two magnetic configurations have different spatial topologies and radial profiles; our aim is to investigate how the presence and strength of the stellar dipole affect the long term evolution of the accretion-ejection system.
Our primary goal is to directly assess how the magnetic field structure regulates angular momentum transport and the launching of outflows, independently of additional physical processes associated with the star-disk interface.

Our paper is organized as follows:

in Section 2, we introduce our model setup,
in Section 3, we present and discuss our simulations,
in Section 4, we compare different runs, focusing on the evolution of mass and angular momentum fluxes and in Section 5, we summarize our main conclusions.

\begin{table}
\centering
\caption{Characteristic parameters of main simulation runs.
The parameters displayed are the initial magnetic field configuration (poloidal disk magnetic field, dipolar  magnetic field or the superposition of both).
The plasma beta $\beta_p= 8 \pi P / B^2$ in all runs is 100,
The simulations listed in the table, correspond to the following runs;
Case1 serves as the reference model.
Case2 and case3 include a purely stellar dipolar magnetic field, representing the weak and strong dipolar strength, respectively.
In cases 4 to 7, the superposition of both the stellar dipole and the poloidal magnetic field is implemented where the strength of the dipolar component increases progressively from Case 4 through Case 7.}


\begin{tabular}{|c|c|c|c|c|c}
\hline
\hline
RunID & B Disk & B Dipole &Vector potential  \\
\hline
\hline
Case 1 &Yes &No &	$A_{\phi,d}$ \\
\hline
Case 2 & No &	Yes &	$10^2A_{\phi,dp}$  \\
\hline
Case 3 & No & Yes &	$10^3A_{\phi,dp}$ \\
\hline
Case 4  &Yes &Yes &$ A_{\phi,dp}+A_{\phi,d}$ \\
\hline
Case 5 &Yes &Yes &$10 A_{\phi,dp}+A_{\phi,d}$ \\
\hline
Case 6 &Yes &Yes &$10^2A_{\phi,dp}+A_{\phi,d}$\\
\hline
Case 7 &Yes &Yes &$10^{3}A_{\phi,dp}+10^{-3}A_{\phi,d}$ \\
\hline
\hline
\end{tabular}
\label{Table:1}
\end{table}

%

\subsection{Our approach and scientific justification}
\label{sec:setup_justification}

Previous numerical studies that include the magnetospheric gap have primarily focused on the physics of star-disk interaction, including magnetospheric accretion, angular momentum exchange, disk locking, accretion funnels, magnetic torques, and reconnection events in the immediate vicinity of the star \citep{2003ApJ...595.1009R,2002ApJ...578..420R,2008A&A...478..155B,Zanni2009}
In contrast, in the present paper we investigate the large scale, steady MHD disk winds launched over a much broader region of the inner part of the accretion disk.
Moreover, \citet{2009ApJ...692..346F} demonstrated that a strong stellar dipolar magnetic field may even decollimate an otherwise established disk wind.
Thus, while magnetospheric simulations are essential for studying the star-disk interface, they address a different physical problem from what is considered here.
Compared to previous works~\citep{Zanni2009,2009ApJ...692..346F,2013A&A...550A..99Z}, our computational domain begins at the inner disk radius; consequently, the inner gap, stellar magnetosphere, and stellar rotation are excluded. This simplification is motivated by several considerations.

First, our study concentrates on large scale jet dynamics, specifically, collimation and stability over distances much larger than the stellar vicinity. These processes are primarily driven by the disk wind rather than by detailed star-disk interaction.
Second, from a computational perspective, including the magnetospheric gap introduces significant complexity \citep{2009ApJ...692..346F}. Resolving the gap requires extremely high resolution to capture the strong gradients in magnetic field strength, density, pressure, and velocity.
By excluding the gap, we conserve computational resources, achieve higher resolution in the jet launching region, and simulate the system for longer dynamical times. This allows the setup to reach a well established steady state.

It is important to clarify that the present work is not a direct comparison between the stellar dipolar component and the disk poloidal field.
The two configurations have different spatial topologies and radial dependencies, and consequently interact with the accretion disk in different ways.
Our aim is instead to investigate how different large scale magnetic field configurations influence the long term evolution of the accretion-ejection system within the radial domain and parameter range considered here.
In particular, we explore how the presence and strength of a stellar dipolar field affect the large scale outflow associated with the disk magnetic field, including its structure, stability, mass flux, angular momentum transport, and collimation.
Our results should therefore not be interpreted as demonstrating that the disk magnetic field is intrinsically more important than the stellar dipole in general.

Our approach is complementary to, but distinct from, the work of \citet{2009ApJ...692..346F}. Both studies consider the interaction between a stellar dipolar magnetic field and a magnetized disk outflow. In Fendt (2009), the initial magnetic configuration consists of a superposition of a stellar dipole and a disk magnetic field, with different relative field orientations considered.
The study investigates the resulting interaction between the stellar and disk outflow components and shows that the stellar magnetic component can substantially modify the disk wind, including its collimation.
In particular, when the stellar contribution becomes sufficiently important, the interaction between the two magnetic components can lead to time dependent and less collimated outflow structures~\citep{2009ApJ...692..346F}.
However, the present study addresses a related but different question.
Rather than focusing primarily on the interaction of the stellar and disk outflow components in the magnetospheric region, we investigate the long term evolution of the accretion-ejection system for different magnetic field configurations and stellar dipole strengths, with emphasis on the conditions under which a persistent and well collimated disk wind is maintained, or suppressed.
In addition, we have included the disk evolution in our setup while ~\citet{2009ApJ...692..346F} prescribed the disk as a boundary condition.
Thus, although both studies address the interaction of stellar and disk magnetic fields, the present work is not intended to reproduce the specific model of Fendt (2009) or to provide a direct comparison of the intrinsic importance of the stellar and disk magnetic components; instead, it is complementary to that work.
Accordingly, the computational domain begins at the inner edge of the accretion disk and does not include the magnetospheric star-disk interaction region.
The stellar dipolar magnetic field is introduced as a prescribed large scale magnetic field threading the computational domain, allowing us to investigate the effects of different magnetic field configurations rather than to establish a direct comparison between the stellar dipole and the disk poloidal field.

Nevertheless, to assess the validity of this simplification and address concerns regarding the exclusion of the inner gap, we performed additional test simulations that explicitly include an inner gap representing the disk truncation radius.
These simulations are listed in Table~\ref{Table:2}, and their results, presented in Appendix~\ref{inner gap runs}, show good agreement with our main simulations.
These simulations were designed as a robustness test of the treatment of the innermost disk region and confirm that our main conclusions are not sensitive to the presence of this gap.

\subsection{Stellar rotation and the propeller regime}
\label{sec:discussion_stellar_rotation}

In the present study, stellar rotation is not included as an additional dynamical parameter because our primary aim is to isolate the effect of the stellar dipole strength and its interaction with the large scale magnetic field threading the disk.

Including stellar rotation would introduce an additional source of magnetic twisting and angular momentum exchange and would require specifying the stellar angular velocity and consistently treating the rotating stellar magnetospheric boundary.
In particular, the differential rotation between the star and the inner disk can twist the stellar magnetic field lines, thereby modifying magnetic stresses and angular momentum exchange between the star and disk, as well as generating flares within the truncation region.

Previous numerical studies have demonstrated the importance of the magnetic coupling between the stellar magnetosphere and the disk in determining the accretion-ejection structure~\citep{1997ApJ...489..890M,2002ApJ...578..420R,2003ApJ...595.1009R,Zanni2009,2013A&A...550A..99Z,2021ApJ...906....4I,2024MNRAS.528.7310M}.
For instance, \citet{1997ApJ...489..890M} showed that rapid stellar rotation can result in a field geometry that inhibits polar accretion even when the magnetic and ram pressures balance at the disk surface.

In addition, \citet{2002ApJ...578..420R} showed that magnetic stresses dominate angular momentum transport between the star and the disk in the quiescent regime, while \citet{2003ApJ...595.1009R} demonstrated that in the propeller regime the magnetosphere expands into a magnetic tower that intermittently blocks accretion, producing quasi-periodic outbursts.
\citet{2013A&A...550A..99Z} showed that magnetospheric ejections driven by the combined rotation of the star and the disk can efficiently extract angular momentum and that the system enters the propeller regime when the disk truncation radius approaches the corotation radius.
\citet{2024MNRAS.528.7310M} found that stellar rotation leads to disk thickening and field line twisting near the truncation region, driving reconnection and the formation of a gap between the star and the disk

In general, a propeller state is expected when the disk truncation radius lies inside the corotation radius, in which case rotation driven outflows could modify the innermost disk region. However, the large scale disk wind studied in this work is launched over a much broader radial range, extending well beyond the truncation region.
The additional test presented in Appendix~B, shows that the large scale jet evolution is insensitive to the details of the innermost region.
This indicates that the launching and collimation of the large scale outflow are governed primarily by the large scale poloidal field threading the disk, rather than by rotation dependent processes in the immediate vicinity of the star.
The omission of stellar rotation is therefore not expected to alter the main conclusions of this work, although it does exclude the magnetospheric and propeller driven phenomena discussed above.

Consequently, a fully self consistent investigation of a rotating stellar magnetosphere would require explicitly including stellar rotation and an appropriate conducting stellar boundary.
Such a treatment is beyond the scope of the present study and is left for future work.

%
\begin{figure*}
\centering
\includegraphics[width=17cm]{\figurepath/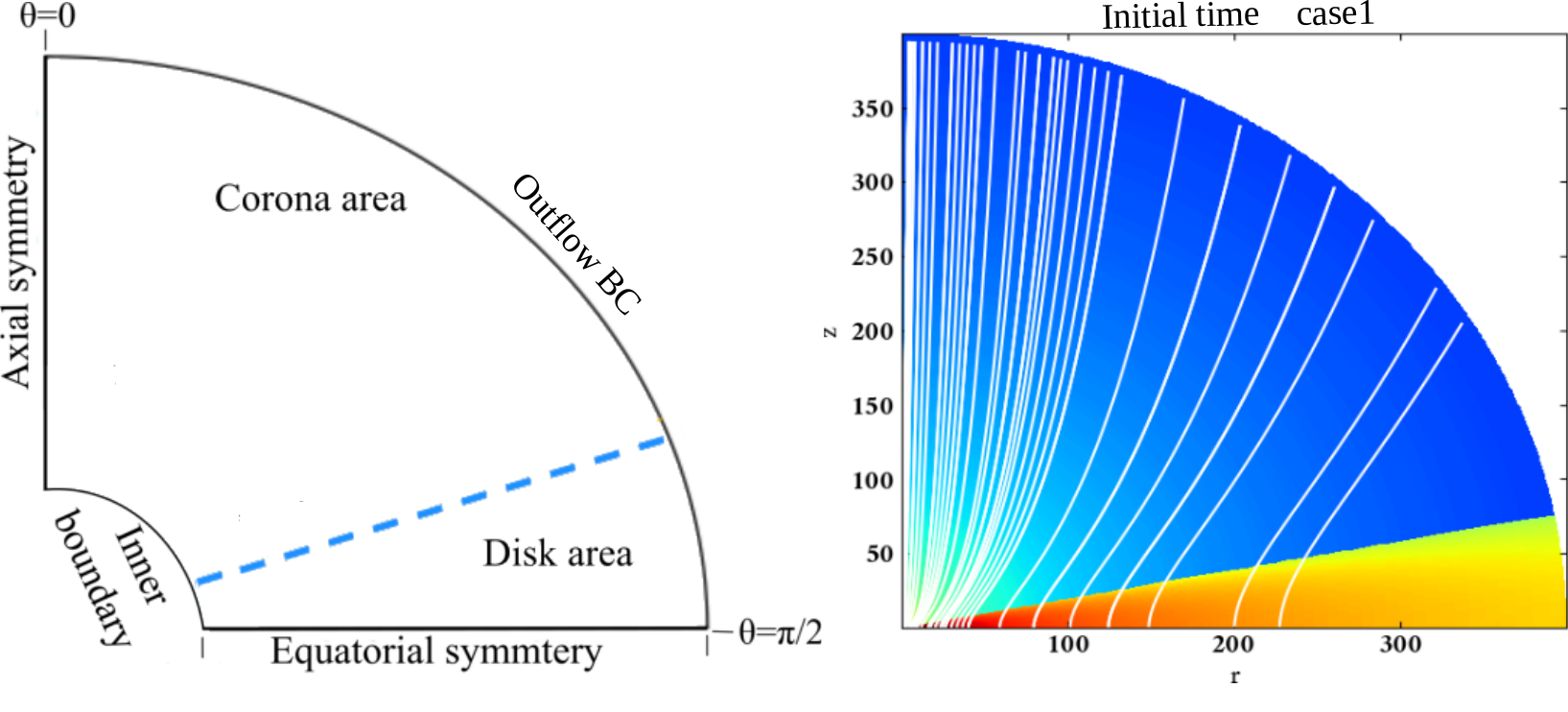}
\caption{The left panel displays the computational domain and the applied boundary conditions for the presented simulations.
The domain is confined between two circles and is defined in spherical coordinates. The dashed line demonstrates the disk surface. The whole computational domain and the initial distribution of the disk and above corona is shown in the right panel.
The domain extends to 400$r_{\rm i}$, and the solid lines display the initial poloidal magnetic field lines.}
\vspace{0.2cm}
\label{computational_domain}
\centering
\includegraphics[width=18cm]{\figurepath/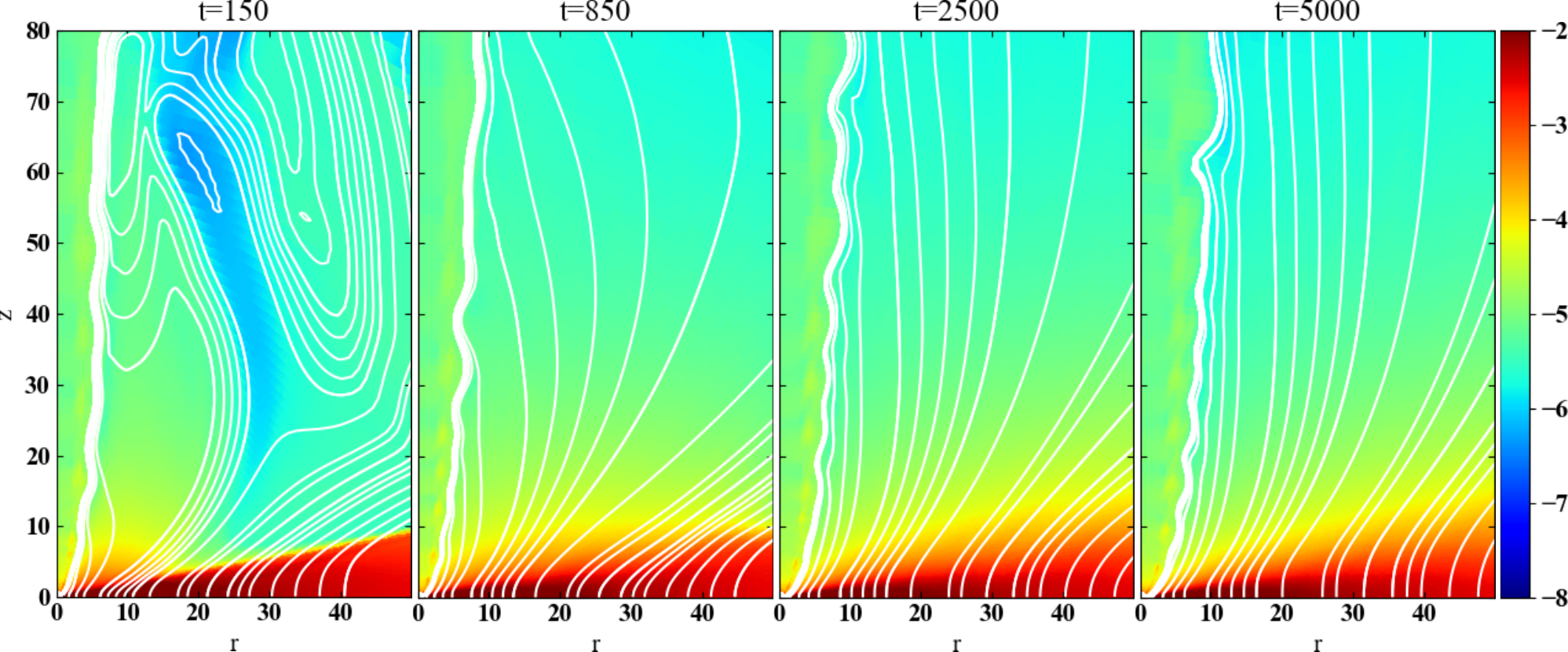}
\caption{Shown are the snapshots of the mass density of the disk-jet system for the reference run (case1) at dynamical times 150, 850, 2500, and 5000.
Solid lines denote magnetic field lines.
The displayed region corresponds to the inner part of the system.
}
\label{rho_reference_run}
\end{figure*}

\section{Model approach}
In this study, we explore the process of jet launching from a magnetised disk under the influence of various magnetic field configurations.
Our primary objective is to examine how the presence of a stellar dipolar magnetic field affects the accretion process and the launching of outflows in protostellar disk-jet systems, such as those associated with classical T Tauri stars.

To analyse the effects of a dipolar magnetic field on the disk-jet system, we first establish an appropriate model setup capable of reaching a steady state.
In this section, we present and discuss the model setup, defining the initial and boundary conditions, as well as the governing equations of the disk-jet system.

\subsection{Governing equations}
We employ the MHD module of the PLUTO code \citep{2007ApJS..170..228M, 2012ApJS..198....7M}, version 4.4.2 to solve the time dependent, resistive, inviscous magnetohydrodynamic (MHD) equations governing the conservation laws for mass, momentum, and energy,
\begin{equation}
\frac{\partial\rho}{\partial t} + \nabla \cdot \left( \rho \vec v \right)=0,
\label{continuity}
\end{equation}
\begin{equation}
\frac{\partial \left( \rho \vec v \right) } {\partial t} +
\nabla \cdot \left(  \rho \vec v \vec v \right) + \nabla P-\frac{ \left( \nabla \times \vec B \right) \times \vec B}{4 \pi}
+ \rho \nabla \Phi = 0.
\label{momentum_eq}
\end{equation}
\begin{multline}
 \frac{\partial e}{\partial t} + \nabla \cdot \left[ \left( e + P + \frac{B^2}{8\pi} \right) \vec v - \left( \vec v \cdot \vec B \right) \frac{\vec B}{4\pi}
 + \left( {\eta} \vec j \right) \times \frac{\vec B}{4\pi} \right]\\
 = - \Lambda_{\rm cool}.
\end{multline}
Here, $\rho$ is the mass density, $\vec v$ is the velocity, $P$ is the thermal gas pressure, $\vec B$ stands for the magnetic field, and $\Phi$ indicates the gravitational potential.
The electric current density $\vec j$ is given by Amp\'ere's law $\vec j = \left( \nabla \times \vec B \right) / 4\pi$ and the total energy density is
\begin{equation}
e = \frac{P}{\gamma - 1} + \frac{\rho v^2}{2} + \frac{B^2}{2} + \rho \Phi.
\end{equation}

We consider an ideal gas with a polytropic equation of state $P = (\gamma - 1) \epsilon$ with  polytropic index of $\gamma = 5/3$ and the internal energy density $\epsilon$.
The cooling term $\Lambda_{\rm cool}$ can be expressed in
terms of Ohmic heating $\Lambda = g\Gamma$, with $\Gamma = \eta {\vec j\cdot \vec j}$, and with
g measuring the fraction of the magnetic energy that is radiated
away instead of being dissipated locally \citep{2007A&A...469..811Z,2012ApJ...757...65S,2014ApJ...793...31S} and
thus here we adopt $g = 1$.

In the above equations, $\eta$ denotes the magnetic diffusivity.
We adopt an $\alpha$-prescription \citep{1973shakuraetal}, consistent with the assumption of a turbulent
origin, as supported by previous studies \citep{2007A&A...469..811Z,2012ApJ...757...65S,2014ApJ...793...31S}.

Following \citealt{2014ApJ...793...31S,2025ApJ...986...51S}, we define the disk diffusivity profile as
$\eta(r,z) = \alpha v_{A}(r,z=0,t) H(r) F_{z,H}$.
Here the Alfv\'en velocity $v_A$ and the disk thermal scale height $H(r) = c_{\rm s}(r,z=0)/\Omega_K(r,z=0)$
are evaluated at the disk mid-plane.
In addition, $\alpha$ is the magnetic diffusivity parameter and its value is 1.6 in our model.
The function $F(z,H)$ further confines the diffusivity to the disk region and up to one scale height above the disk surface, and it is defined as

\vspace{0.2cm}
$ F{(z,H(r))} = \left\{
    \begin{array}{lll}
       \exp\left(-0.5\left(\frac{z-H(r)}{H(r)}\right)^{2} \right) & &  z>H(r)\\
       1                            &  &   z\le H(r)\\
    \end{array}
\right.$
\vspace{0.3cm}

The evolution of the magnetic field is described by the induction equation,
\begin{equation}
\frac{\partial \vec B}{\partial t} - \nabla\times \left( \vec v \times \vec B - \eta \vec j \right) = 0.
\end{equation}

\subsection{Numerical setup}
In this section, we provide a comprehensive overview of the initial and boundary conditions implemented in our model.

Figure \ref{computational_domain} demonstrates the computational domain along with the corresponding boundary conditions.
The domain spans an angle range of $0$ to $\pi/2$ in the poloidal direction, with a total of 256 uniformly spaced grid cells.
In the radial direction, the grid extends over a distance of $0$ to $400 r_{\rm i}$ and consists of 256 stretched grid cells.
\subsection{Boundary conditions}
\label{section:BC}

The boundary conditions are mostly adopted from those used in \citet{2014ApJ...793...31S,2014ApJ...796...29S}.
At the polar axis, axisymmetric boundary conditions are applied. Therefore, the variables do not depend on the azimuthal angle $\phi$.
At the outer radial boundary of the computational domain, the standard outflow boundary conditions of the PLUTO code are applied, where the variables are directly copied into the ghost cells.
At the inner radial boundary of the computational domain, located at $R=1$, we apply a composite boundary condition that distinguishes between the disk and corona regions, which we refer to as the ``inner boundary''.
Across this inner boundary, the disk region is defined by $\theta \geq \pi/2-\epsilon$, where $\epsilon$ denotes the disk scale height.
We emphasize that the present work considers only the upper hemisphere, and the physical conditions of the disk are imposed only within this region. The disk is initialized in hydrostatic equilibrium with the overlying corona.
At the disk mid-plane ($\theta=\pi/2$), we apply the standard ``eqtsymmetric'' boundary condition \citep{2014ApJ...793...31S} provided by the PLUTO code.
This condition enforces equatorial symmetry with respect to the mid-plane. It is similar to a reflective boundary condition, except that the normal component of the magnetic field changes sign across the equatorial plane, while the tangential magnetic field components remain continuous, with the corresponding velocity components transformed consistently.
This condition is specifically designed for simulations that assume symmetry about the equatorial plane and is therefore fully consistent with our model setup.

Across the inner boundary and in the disk region, the pressure, $P$, and density, $\rho$, are prescribed by radial power law profiles.
For the radial velocity $v_{r}$, an accretion condition is applied (not allowing for the entering the material into the domain), and the polar velocity $v_\theta$ is copied into the Ghost cells.
The azimuthal velocity $v_{\phi}$ is prescribed by conserving the specific angular momentum, while the toroidal magnetic field is set by assuming conservation of the toroidal magnetic flux, $R B_{\phi}=\mathrm{const.}$
The remaining magnetic field components are copied into the ghost cells.

In addition, across the inner boundary but for the corona region, $\theta < \pi/2-\epsilon$,  above the disk, the initial density and pressure of the corona are used. To define the radial and polar velocity components, we impose the ideal MHD condition $E_{\phi}=0$, which ensures $v_p \parallel B_p$.
A weak inflow is prescribed along the coronal boundary with $v_{\rm p}=0.02$ to avoid the development of excessively low densities in the coronal region.
We also clarify that $\vec v_{\rm p}$ denotes the poloidal velocity, i.e., the component of the velocity in the meridional plane. It can be expressed in cylindrical and spherical coordinates as

\begin{equation}
\vec{v}_{\rm p}
=
v_r\,\hat{\mathbf{r}}
+
v_z\,\hat{\mathbf{z}}
=
v_R\,\hat{\mathbf{R}}
+
v_\theta\,\hat{\boldsymbol{\theta}},
\end{equation}

where $(r,z)$ represent the cylindrical coordinates, while $(R,\theta)$ denote the spherical coordinates adopted in our simulations.
To achieve a more stable structure, the azimuthal components of the velocity and magnetic field in the coronal region, across the inner boundary are set to zero.

\subsection{Initial Conditions}
Here, we explain the initial conditions of the disk-jet system defined in spherical coordinates.
In our model setup, the accretion disk is initially in hydrostatic equilibrium with a non-rotating corona above it.
We define a thin disk with $\epsilon = H(r)/r = 0.1$.

Our approach builds on previous studies of jet launching from disks \citep{2007A&A...469..811Z,2012ApJ...757...65S} and specifically follows the methodology outlined in \citet{2014ApJ...793...31S}.
Accordingly, the initial density and pressure profiles of the disk are given by:

\begin{align}
\rho_d=\rho_{\rm d,i}\left[\frac{2}{5H^2}\left(\frac{1}{R}-\frac{(1-2.5\epsilon^2)}{Rsin(\theta)}\right)\right]^{1.5},
\label{rho_d} 
\end{align}

\begin{align}
p_d=p_{\rm d,i}\left(\rho_d/\rho_{\rm d,i}\right)^{5/3}.
\label{p_d}
\end{align}
Here, $r$ and $R$ correspond to the cylindrical and the spherical radius, respectively.
$\rho_{\rm d,i}$ and $p_{\rm d, i}$ represent the gas density and pressure at the inner radius of the disk $r_{\rm i}$, respectively.

The initial poloidal velocity of the gas is set to zero ($v_{\theta}=0$), and the gas is prescribed in  sub-Keplerian rotation speed to maintain radial equilibrium in the disk.
The Azimuthal and radial velocity of the gas in the disk are given by,
 \begin{align}
v_{\phi,\rm d}=\sqrt{1-\frac{5\epsilon^2}{2}}\sqrt{\frac{GM}{r}},
\label{v_phi_d}
\end{align}
\begin{align}
v_{\rm R,d}=-sin(\theta)\sqrt{2\mu}\epsilon^2 r^{-1/2} \left(1.25+\frac{1.666}{m^2}\right).
\label{v_r_d}
\end{align}
The density contrast between the gas in the disk and the corona above the disk is denoted as $\delta = 10^{-3}$.
Furthermore, the gas density and the pressure in the corona are,
\begin{align}
\rho_{\rm a}=\rho_{\rm a,i}\left(\frac{r_i}{R}\right)^{(1/\gamma-1)},
\label{rho_a}
\end{align}
\begin{align}
p_{\rm a}=\rho_{\rm a,i}\left(\frac{\gamma-1}{\gamma}\right)\left(\frac{GM}{r_i}\right) \left(\frac {r_i}{R}\right)^{(\gamma/\gamma-1)}.
\label{p_a}
\end{align}
%


\begin{figure}
\centering
\includegraphics[width=1.0\columnwidth]{\figurepath/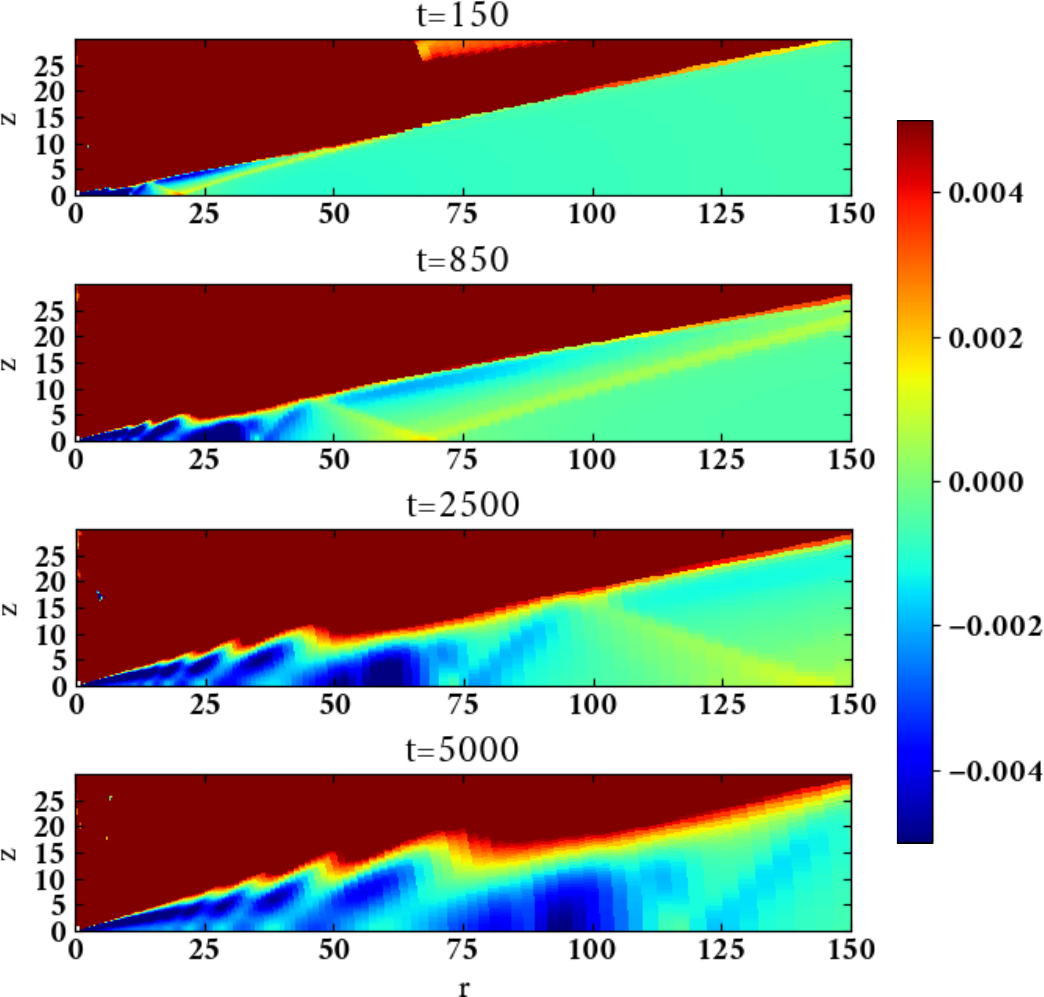}
\caption{Shown are the snapshots of the radial velocity in the disk, for the reference run (Case1) at  dynamical times 150, 850, 2500, 5000. The displayed region corresponds to the innermost part of the system.
}
\label{vr_reference_run}
\end{figure}
\subsection{Initial Magnetic field}
It is now generally accepted that the magnetic field in a protostellar disk-jet system consists of two components, a central stellar dipolar field and an additional component anchored in the circumstellar disk~\citep{2009ApJ...692..346F,2009A&A...502..217M}.

Our primary aim is to investigate how the magnetic field configuration regulates angular momentum transport and the launching of outflows in disk-jet systems.
Accordingly, we consider three magnetic field configurations:

(i) an initially large scale poloidal magnetic field threading the disk, (ii) a purely stellar dipolar magnetic field, and (iii) the superposition of both components.
We therefore consider two distinct magnetic field geometries that govern the disk-jet system.
The first configuration is the initial poloidal magnetic field, following \citet{2007A&A...469..811Z,2012EAS....58..113S,2014ApJ...793...31S}, prescribed by the magnetic flux function $\Psi$:

\begin{align}
\Psi_{\rm d}=\frac{4}{3}B_{\rm i} r^2
\left(\frac{r}{r_{\rm i}}\right)^{3/4}
\frac{m^{5/4}}
{\left[m^2+\left(\frac{z}{r}\right)^2\right]^{5/8}}.
\label{Psi_disk}
\end{align}

The vector potential is calculated by

\begin{align}
A_{\phi,\rm d} = \frac{\Psi_{\rm d}}{R\sin\theta}
\label{A_disk}
\end{align}

where, the parameter $m$ specifies the bending of magnetic field lines and is set to $m=0.6$ in all presented simulations.

Additionally, we employ an alternative magnetic field geometry generated by the stellar dipole.
Following \citet{2000A&A...363..208F,2009A&A...508.1117Z,2009ApJ...692..346F}, the stellar dipole magnetic field, generated by the central protostar, is defined by the following vector potential:
\begin{align}
A_{\phi,{\rm dp}}=B_{\star} R_{\star}^3 \frac{\sin^3\theta}{r^2}.
\label{A_dipole}
\end{align}

Here, $B_{\star}$ denotes the strength of the stellar dipolar field, ranging from $1$ to $1000 \, B_{\rm i}$, where $B_{\rm i}$ is the magnetic field normalization (see Section~\ref{units}).
$R_{\star}$ is the stellar radius and is approximately $1/5$ of the disk inner radius $r_{\rm i}$.
The components of the magnetic field are defined as
\[
B_{\rm R}=\frac{1}{r\sin\theta}\pdv{(A_{\rm \phi}\sin\theta)}{\theta}, \qquad
B_{\rm \theta}=\frac{1}{r}\pdv{(rA_{\rm \phi})}{r}.
\]

\begin{figure*}
\centering
\includegraphics[width=15.5cm]{\figurepath/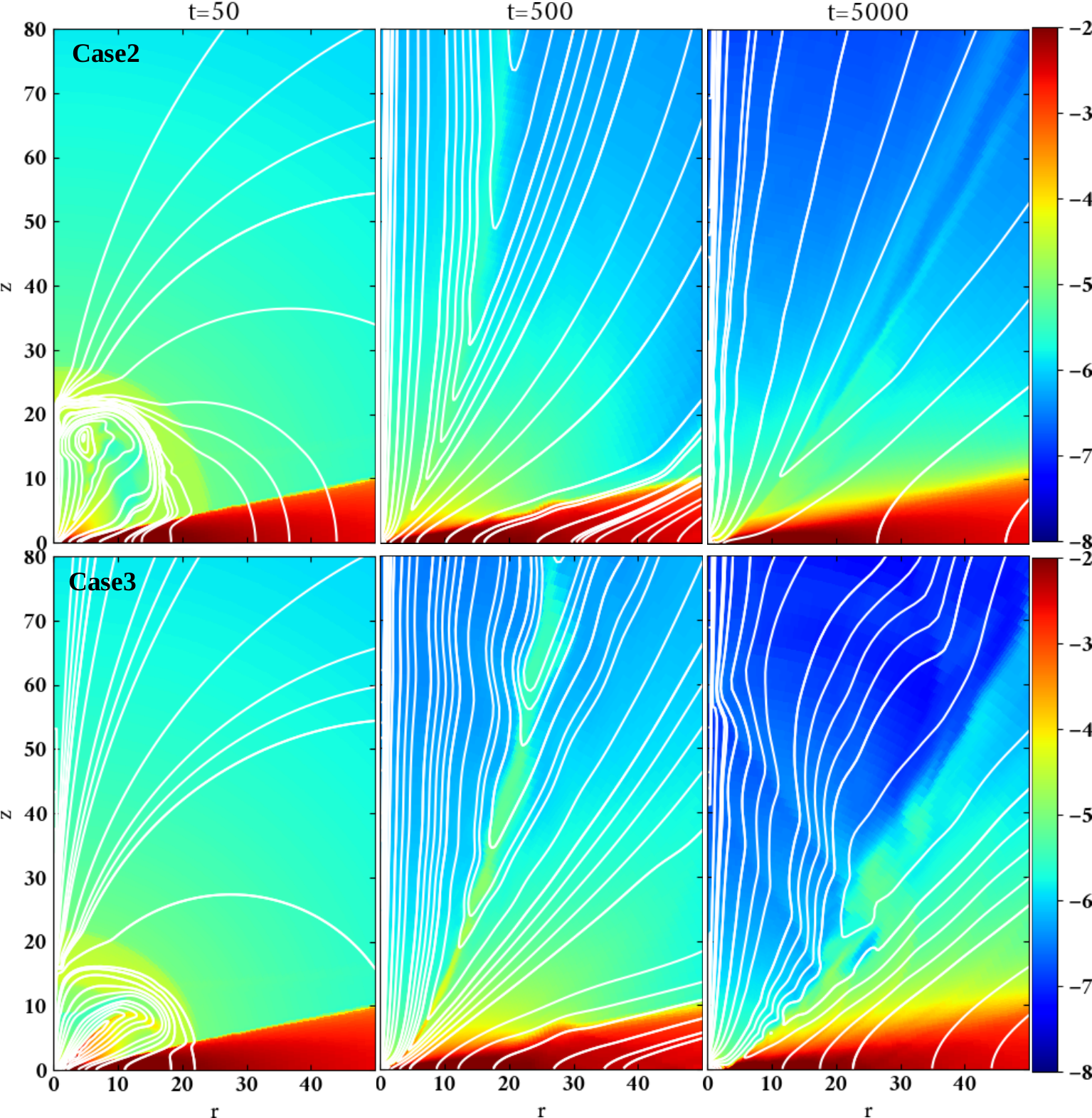}
\caption{Shown are snapshots of the mass density distribution for runs that include only the stellar dipolar magnetic field (Case 2 and Case 3). Case 3 corresponds to a stronger field ($B_\star = 1000$), while Case 2 represents a weaker field ($B_\star = 100$). The snapshots are taken at dynamical times $t = 50$, $500$, and $5000$.}
\label{density_Pure_dipole}
\end{figure*}

\subsection{Units and Normalization}
\label{units}
In this section, we describe the normalization employed throughout the paper.
All spatial quantities are expressed in units of the inner disk radius denoted by $r_{\rm i}$, which is set to unity $r_{\rm i}=1$.
The Keplerian velocity at the inner radius of the disk is $v_{\rm k, i}$  which is set to unity and is used as the unit of velocity.
Time is estimated in units of $t_{\rm i}=\frac{r_{\rm i}}{v_{\rm k,i}}$.
The gas density and pressure in the disk are normalized to their values at the inner disk radius defined as $\rho_{\rm d, i}$ and $p_{\rm d, i}=\epsilon^2\rho_{\rm d,i}v_k^2$.
The magnetic field is measured in units of the magnetic field at the inner disk radius $B_{\rm i}$ which is defined by,
$B_{\rm i}=\sqrt{\frac{2 P_{\rm d,i} }{\beta_{\rm p} } }$.
The $\beta$ parameter denotes the ratio of the thermal gas pressure to the magnetic pressure evaluated at this radius, and $\beta$ is $100$ in all runs\footnote{In PLUTO code the magnetic field is normalized considering $4\pi =1$ \label{MyFootNoteLabel}}.
\section{Results and discussion}
\label{sec:results_main_runs}
In this section, we present the simulation results, for the various magnetic field configurations that launch the magnetized jet, as listed in Table~\ref{Table:1}.
The simulations include case 1 as the reference run, cases 2 and 3 which include a purely stellar dipole field, corresponding to weak and strong dipolar configurations, respectively.
Cases 4 to 7 incorporate the superposition of both the stellar dipolar and the poloidal magnetic field, with the dipolar component increasing progressively across these cases.

A comparative plot of the initial magnetic field strengths in the inner region of the disk is presented in Figure~\ref{Bp_strength_comp1}. It illustrates the radial profiles of the dipolar stellar magnetic field and the poloidal disk magnetic field at the disk mid-plane.
These profiles correspond to simulation cases 4 to 7 as listed in Table~\ref{Table:1}, which include the superposition of both magnetic field configurations.

First, we present our reference run and discuss its characteristics.
Subsequently, we implement a stellar dipole magnetic field and investigate the resulting differences.
For the visualization and analysis of the simulation data, we use the latest version of the PyPLUTO package \citep{2025JOSS...10.8448M}.
\begin{figure*}
\centering
\includegraphics[width=18cm]{\figurepath/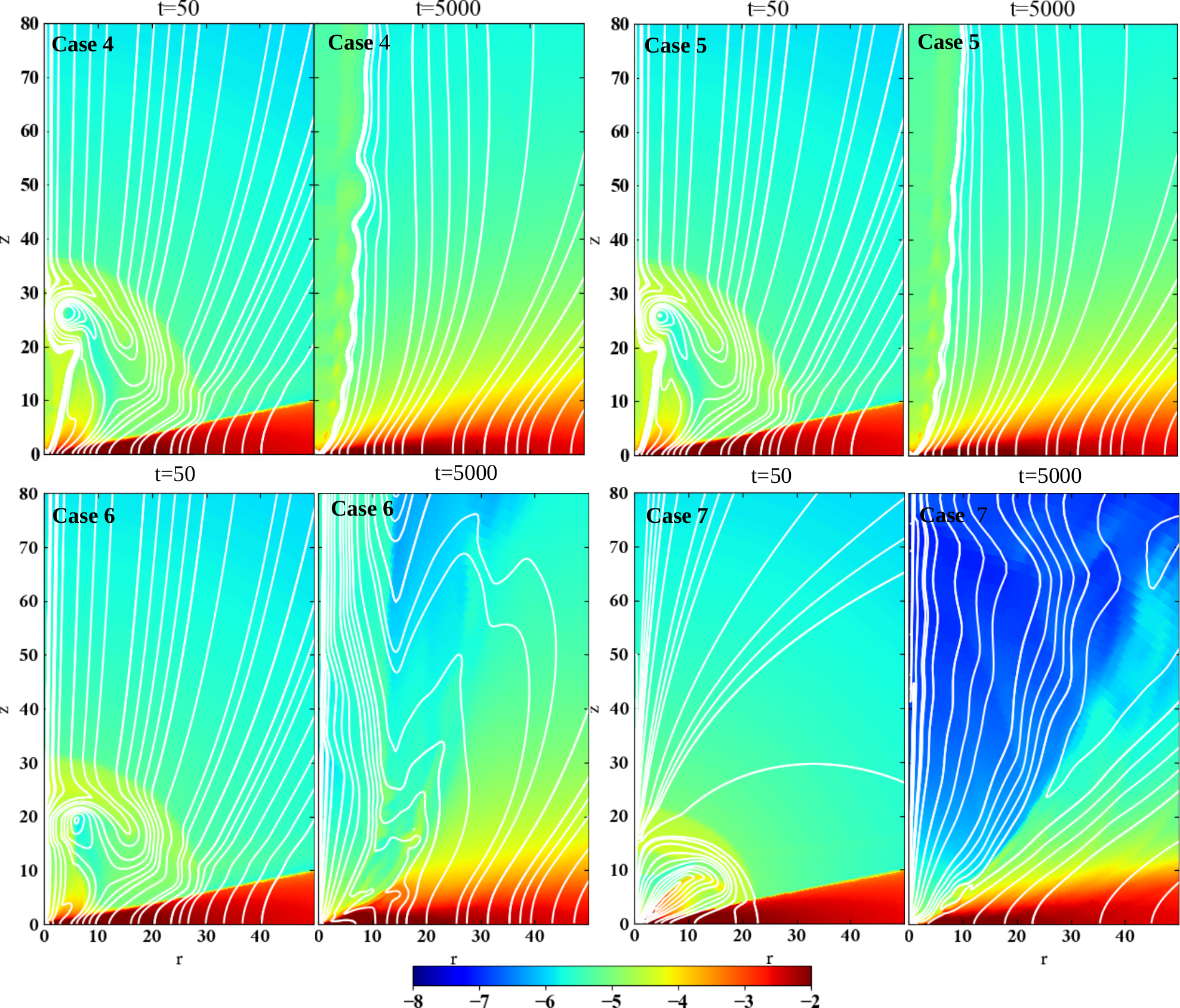}
\caption{Shown are snapshots of the mass density maps for simulations comprising the superposition of the poloidal and stellar dipolar magnetic field, in logarithmic scale and at dynamical times $t=50, 5000$, for cases 4 to 7 as listed in Table~\ref{Table:1}.}
\label{rho_dipole_and_disk}
\end{figure*}
\begin{figure*}
\centering
\includegraphics[width=18cm]{\figurepath/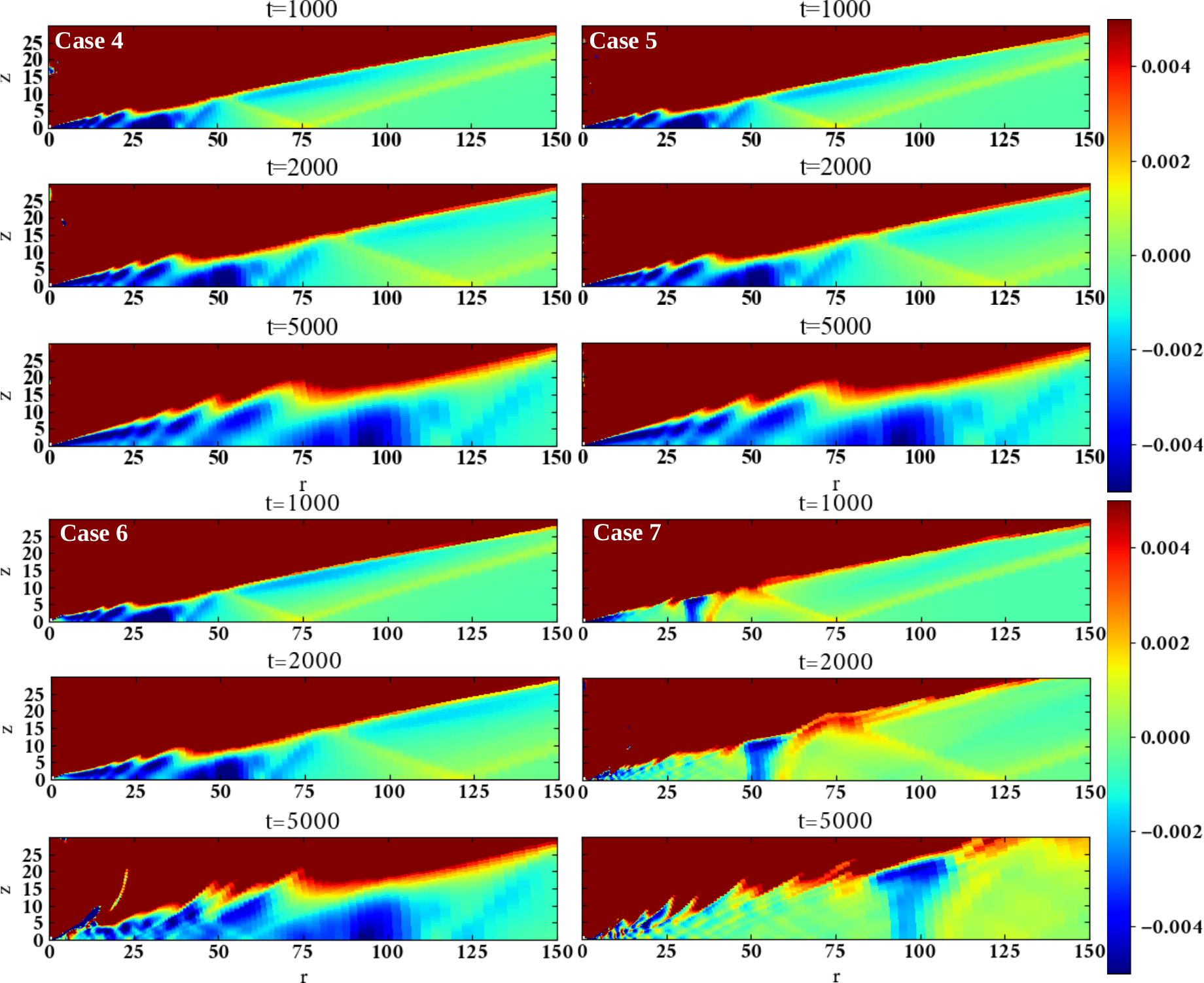}
\caption{Shown are snapshots of the radial velocity of the disk region, for simulations incorporating the superposition of the the poloidal and the stellar dipolar magnetic field, corresponding to cases 4 to 7 in Table~\ref{Table:1}, at dynamical times, $t=1000, 2000, 5000$.}
\label{vr_dipole_and_disk}
\end{figure*}
\subsection{Reference run with initial poloidal magnetic Field}
We begin by presenting the simulation results incorporating the initial poloidal magnetic field defined by Eq.~\ref{Psi_disk}.
This serves as our reference run and referred to as case 1 in Table~\ref{Table:1}.
All simulations were run for 5000~$t_{\rm i}$ dynamical times, corresponding to 800 revolutions at the inner disk radius $r_{\rm i}$.

Figure~\ref{rho_reference_run} shows the evolution of the mass density of the disk and the formed jet during time corresponding to the inner region of the system.
From the figure, it is observed that a stable and continuous outflow is launched from the inner regions of the disk and expands to the larger distances.
A well collimated jet is formed at the inner part of the disk.
During the early stages, the interaction between the jet and the surrounding environment leads to the formation of a bow shock.
As the system evolves, the outflow expands further, eventually reaching larger scales and leaving the computational domain.

The formation of the jet is based on the MHD disk wind model proposed by \citet{1982MNRAS.199..883B,1983ApJ...274..677P}.
In this framework, the jet material is predominantly launched from the accretion disk.
The accretion process plays a crucial role and we can examine this process by considering the radial velocity map of the disk.
Figure~\ref{vr_reference_run} displays snapshots of the radial velocity from the reference run, case 1, at different dynamical times.
It is seen that accretion is initially established in the inner regions of the disk and extends to larger radii over time.
This behavior is primarily driven by the magnetic torque exerted on the disk, which efficiently extracts angular momentum and thus facilitates the inward flow of material.

It should be noted that the reference simulation was run to 20000~$t_{\rm i}$ dynamical times demonstrating the long term stability of the reference setup and can be reliably used for further investigation.

\subsection{Implementing the stellar dipolar magnetic field}
In this section, we present and discuss the results of
simulations threaded by purely weak and strong stellar dipolar magnetic fields, corresponding to case 2 and case 3 in Table~\ref{Table:1}, respectively.

\subsubsection{Pure stellar dipole magnetic field}
\label{pure dipole}
The stellar dipolar magnetic field has been implemented in runs Case 2 and Case 3 (see Table~\ref{Table:1}).
In these simulations, the disk-jet system is threaded
exclusively by a stellar dipolar field prescribed by Eq.~\ref{A_dipole}.
The dipolar field strength is parameterized by $B_\star$.
It should be noted that the stellar surface is not included in our simulation domain and the dipolar magnetic field experienced by the disk, both at the inner disk radius and farther out, is significantly weaker than the field strength at the stellar surface.

Figure~\ref{density_Pure_dipole} demonstrates the density evolution of the disk-jet system in Case 2 and Case 3.
Case 3 corresponds to a stronger magnetic field ($B_\star\,=\,1000$), while Case 2 represents a weaker magnetic field ($B_\star\,=\,100$).
The snapshots are taken at dynamical times  $t= 50, 500$, and $ 5000$.
From the figure, we find the following results:

It is evident that the ejected material does not resemble the typical outflow observed in the reference run, highlighting the importance of the magnetic field configuration, as expected from MHD disk wind theory.
In fact, the magnetic field lines show less collimation than the reference run, and only a small amount of mass transfer from the disk is observed.
As a result, a purely dipolar magnetic field is unable to efficiently remove angular momentum from the disk, and consequently, no significant accretion occurs.
Furthermore, the outflow lacks proper collimation and acceleration above the disk.
In the disk region, the magnetic torque remains too weak to efficiently extract angular momentum and to drive a well structured and collimated outflow from the disk
(see Section~4).

These findings are consistent with the results discussed by \citet{2000A&A...363..208F}. They investigated a dipolar stellar magnetosphere interacting with a Keplerian disk (treated as a boundary) and demonstrated that the emerging two-component outflow from the star and disk remained largely uncollimated. This behavior was attributed to the absence of a net electric current along the outflow, a consequence of the initial dipolar field configuration.

Furthermore, some fluctuations appear in the outflow and along the field lines in Case 3, which corresponds to the stronger dipolar magnetic field.
These fluctuations seem to be caused by the perturbed structure in the underlying disk.

The magnitude of the stellar dipole is an important parameter in determining the extent to which the stellar magnetosphere influences the accretion-ejection structure.
In runs with pure dipolar field, increasing $B_*$ modifies the magnetospheric structure and the associated outflow.
Nevertheless, within the present setup, the stellar dipole alone does not produce the persistent and well collimated outflow obtained in the reference case, in which a large scale poloidal magnetic field threads the disk.
Thus, while the dipole strength determines the extent and dynamical influence of the stellar magnetosphere, our results indicate that the large scale  magnetic field threading disk is essential for maintaining the well collimated outflow found in the reference model.
We note that a more complete treatment of the star-disk interaction, including a self-consistent rotating magnetosphere, could introduce additional magnetic twisting and outflow components beyond the disk truncation region.

\begin{figure}
\centering
\includegraphics[width=0.9\columnwidth]{\figurepath/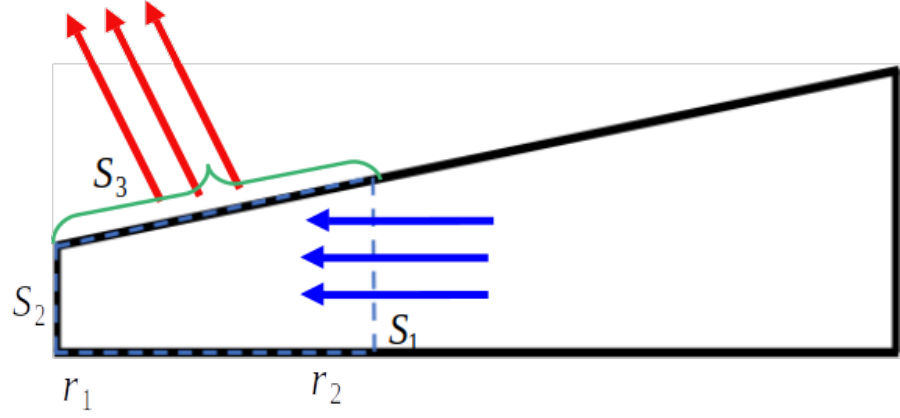}
\caption{Control volume used to measure the accretion and ejection mass fluxes within the disk-jet structure.}
\label{control_volume}
\end{figure}

\subsubsection{Superposition of stellar dipole and disk magnetic fields}
In this section, we present a series of simulations in which a superposition of the initial poloidal magnetic field and the stellar dipolar magnetic field was implemented, namely, cases 4, 5, 6, and 7, corresponding to dipolar magnetic field  strengths of $B_{\star}\,=\,1, 10, 100$, and $1000$, respectively, as listed in Table~\ref{Table:1}.
In this sequence of runs, the influence of the stellar dipolar magnetic field gradually increases, while the relative contribution from the disk magnetic field correspondingly decreases.
In particular, in Case 7, the poloidal magnetic field strength of the disk was further reduced by a factor of 1000 to make the dipolar component even more dominant, resulting in a very weak poloidal disk magnetic field, i.e., $B_{\star} + 0.001 B_{\rm disk}$.

It should be emphasized that we do not investigate different relative alignments between the stellar dipolar magnetic field and the poloidal disk magnetic field as discussed by \citet{2009ApJ...692..346F}.
In our simulations, the stellar dipolar and disk poloidal magnetic fields have the same direction at the northern disk surface; accordingly, the two fields are aligned in this region.
The X-point location depends on the relative strength of the two fields and can differ among runs, but it is not located inside the disk.
In addition, the disk field is prescribed by Eq.~\ref{A_disk} as defined by \citet{2007A&A...469..811Z}, while \citet{2009ApJ...692..346F} applied the model of \citet{1997ApJ...482..712O,2002A&A...395.1045F} for the disk field component.
The density distributions of cases 4 to 7 are shown in Figure~\ref{rho_dipole_and_disk}.

A clear trend emerges from these simulations when the disk poloidal magnetic field contribution is dominant, the system develops a well structured outflow that extends to larger distances.
For instance, in cases 4 and 5, the outflows remain smooth and exhibit good collimation in the inner outflow region.
However, as the strength of the stellar dipolar magnetic field increases, perturbations and instabilities begin to appear inside the outflow along the field lines. 
For instance, in Case 6, where the dipolar field contribution becomes significant, the resulting outflow is less smooth.
This effect becomes even more pronounced in Case 7, where the dipolar field dominates.
The ejected material no longer exhibits the collimation as seen in the reference run.
This finding is in agreement with the results obtained by \citet{2009ApJ...692..346F}.
They also found that strong stellar winds lead to low collimation in the outflow.
These results further support the conclusion that
well collimated protostellar jets are primarily driven by disk winds rather than the stellar magnetosphere.

Furthermore, the magnetic field lines exhibit fluctuations, reflecting the perturbation in the underlying disk and within the outflow.
Additionally, the radial velocity distributions for Case 4 through Case 7 are shown in Figure~\ref{vr_dipole_and_disk}.
In the simulations where the disk magnetic field dominates (cases 4 and 5), a well defined accretion flow develops and extends smoothly to larger radii.
In Case 6, where the stellar dipolar magnetic field is stronger, the accretion flow begins to exhibit enhanced perturbations, particularly in the inner disk region.

In Case 7 characterized by a significantly stronger dipolar magnetic field, the accretion flow is substantially weakened.
Small amount of accretion occurs in the inner part of the disk, while the disk region is strongly perturbed.
These results further confirm that angular momentum removal becomes inefficient when the stellar dipolar magnetic field is dominant within the disk.

In summary, these simulations demonstrate that the relative strength of the stellar dipolar and poloidal magnetic fields critically governs the evolution of the disk and the resulting jet.

When the disk magnetic field is sufficiently strong (as in Cases 4 and 5), angular momentum is efficiently removed, driving smooth, well collimated outflows, and sustained accretion across the disk is established.
Conversely, when the stellar dipolar magnetic field becomes significant or dominant (as in Cases 6 and 7), accretion is disrupted. In addition, fluctuations appear within the disk and outflow, and the outflow becomes uncollimated.

These results highlight the tight coupling between disk accretion and jet evolution, showing that inefficient accretion in the disk leads to disrupted outflows.

\begin{figure}
\centering
\includegraphics[width=1.0\columnwidth]{\figurepath/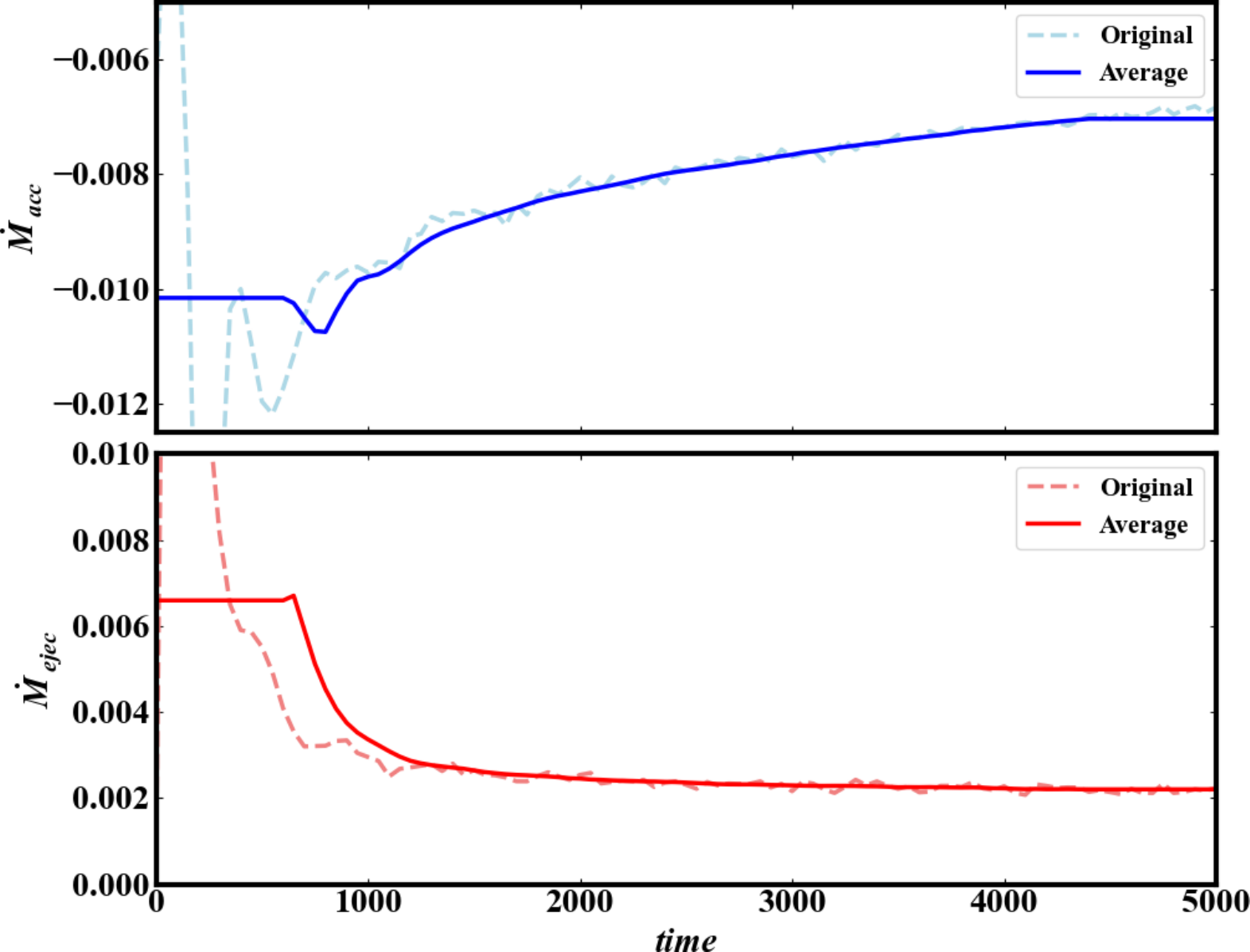}
\caption{Shown is the time evolution of the accretion mass flux $\dot M_{\rm acc}$ and the outflow mass flux $\dot M_{\rm ejec}$ for the reference run (Case 1), evaluated within the control volume defined in Figure~\ref{control_volume}.
The solid lines represent the time averaged mass fluxes, while the dashed lines show the raw (original) values.}
\label{flux_case1}
\end{figure}

\begin{figure*}
\centering
\includegraphics[width=18cm]{\figurepath/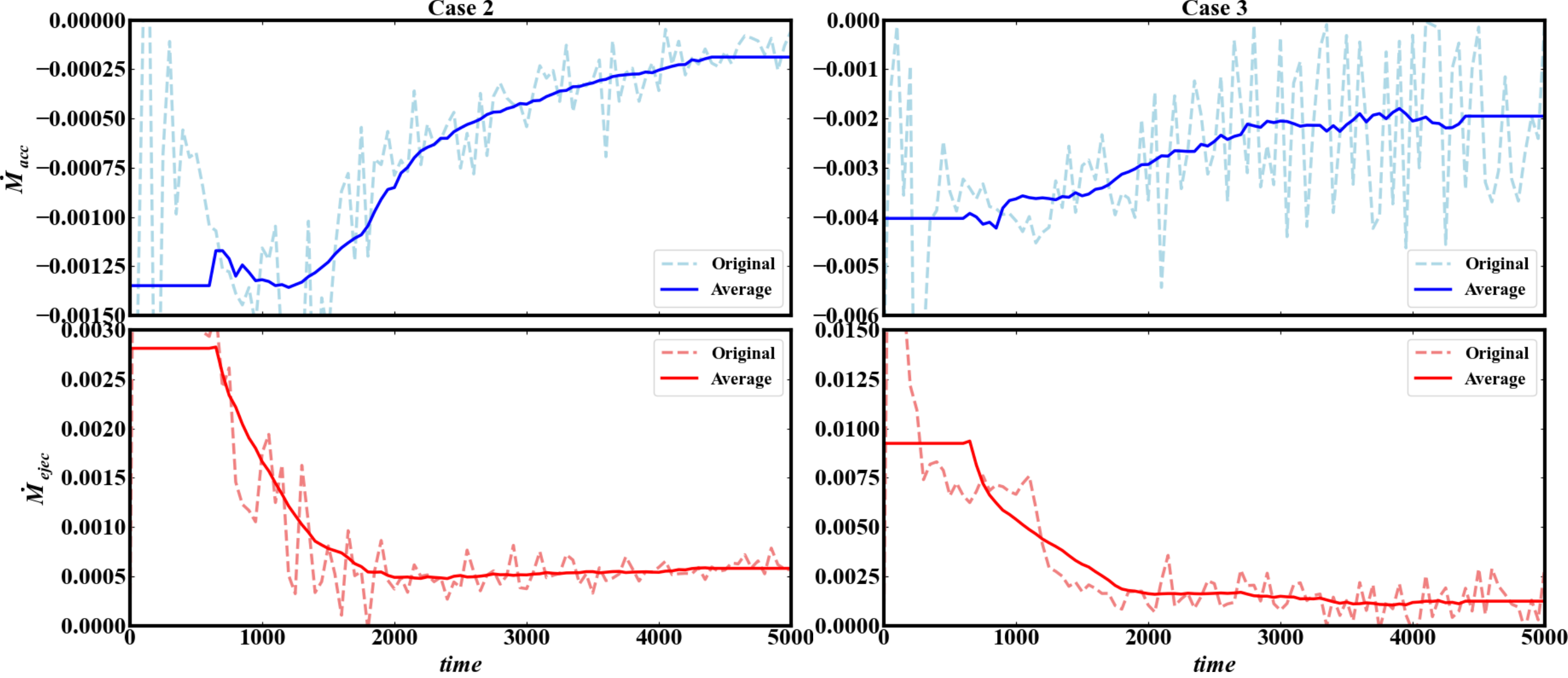}
\caption{Shown is the time evolution of the accretion mass flux $\dot M_{\rm acc}$ and the outflow mass flux $\dot M_{\rm ejec}$ for Cases 2 and 3, corresponding to the weak and strong dipolar magnetic fields, respectively, evaluated within the control volume defined in Figure~\ref{control_volume}.
The solid lines represent the time averaged mass fluxes, while the dashed lines show the raw (original) values.}
\label{flux_case2_and_case3}
\end{figure*}
\begin{figure*}
 \centering
 \vspace*{0.1 cm}
\includegraphics[width=18cm]{\figurepath/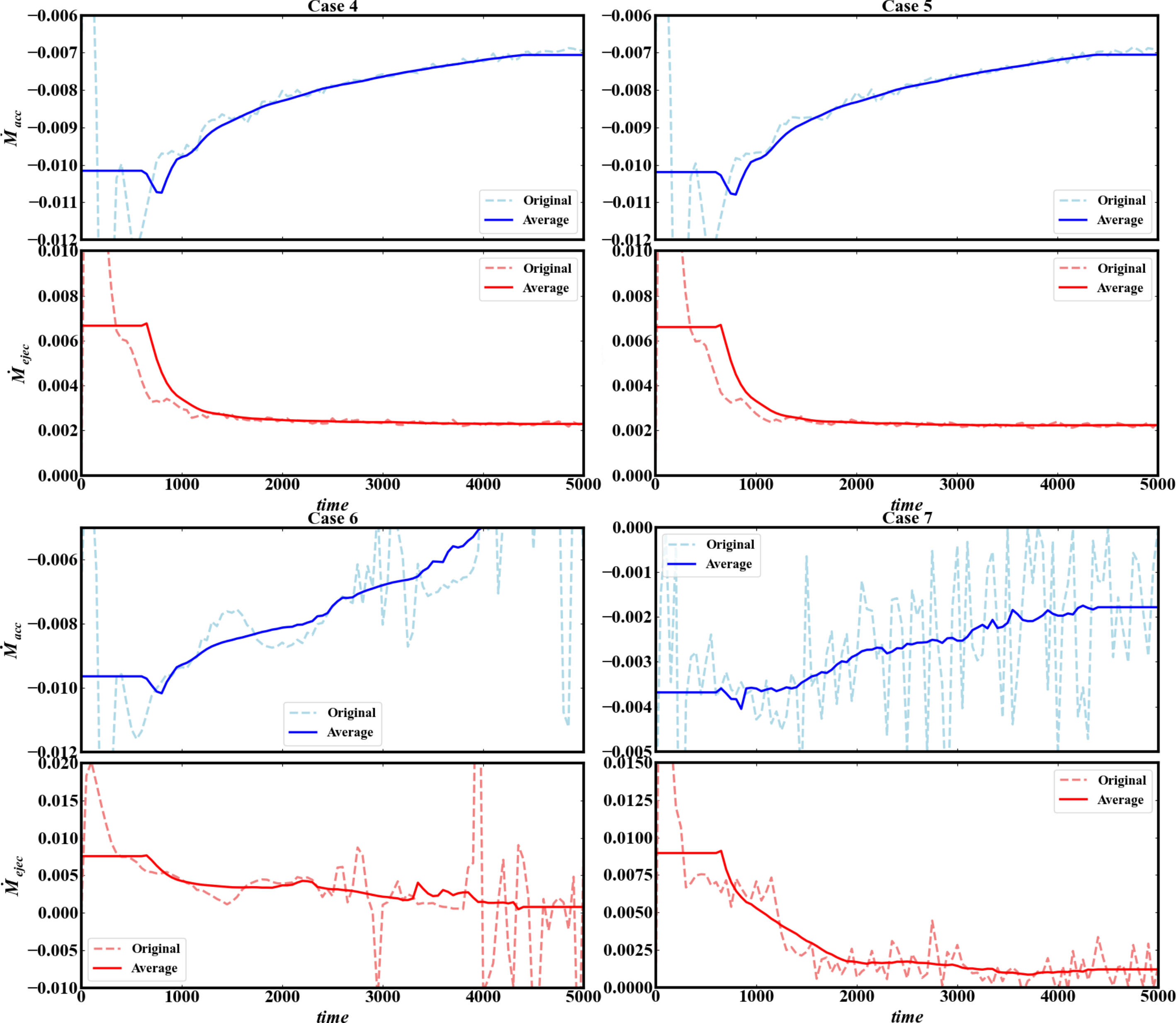}
\caption{Shown are the accretion mass flux ($\dot M_{\mathrm{acc}}$) and the outflow mass flux ($\dot M_{\mathrm{ejec}}$) for cases 4 to 7 (see Table~\ref{Table:1}), which include the superposition of both magnetic fields.}
\label{flux_case4_to_case7}
\end{figure*}

\begin{figure}
\centering
\includegraphics[width=1.0\columnwidth]{\figurepath/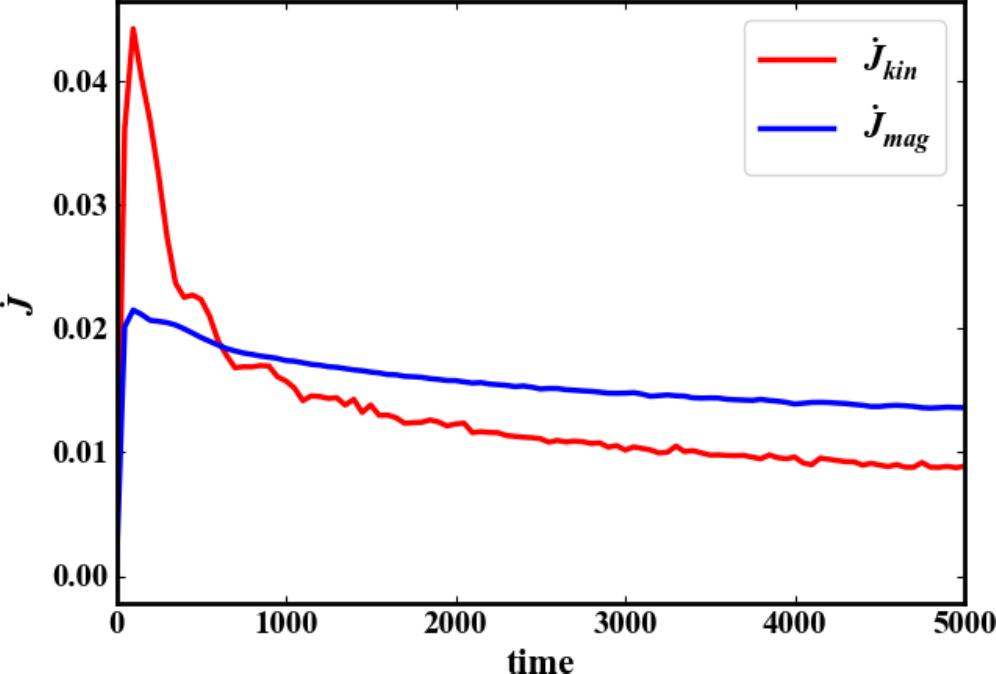}
\caption{Shown is the time evolution of the kinetic ($\dot J_{\rm kin}$) and magnetic ($\dot J_{\rm mag}$) angular momentum fluxes carried by the outflow for the reference run. These fluxes are defined by Equations~\ref{Torque1} and~\ref{Torque2}.}
\label{Torque_ref}
\end{figure}
\begin{figure*}
\centering
\includegraphics[width=18cm]{\figurepath/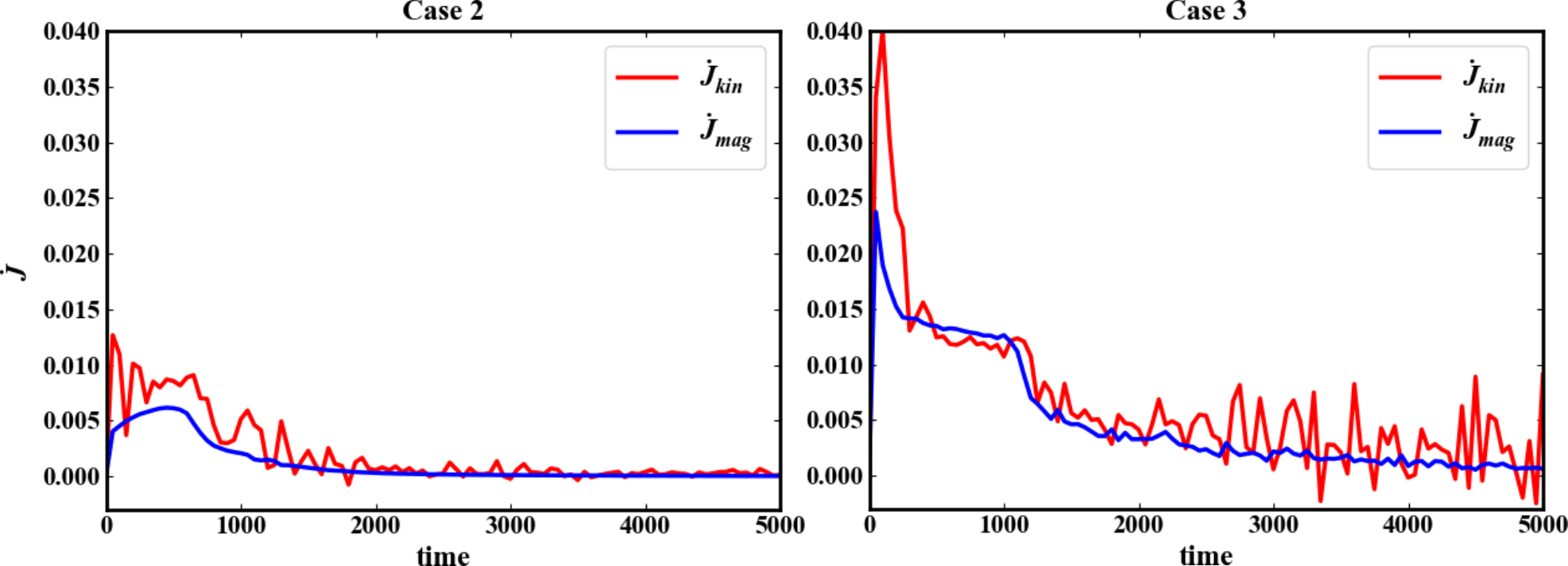}
\caption{Shown are the time evolution of the kinetic $\dot J_{\rm kin}$ and magnetic $\dot J_{\rm mag}$ angular momentum fluxes, carried by the outflow and are given by Equations \ref{Torque1} and \ref{Torque2}, for Case 2 (left panels) and Case 3 (right panels) including the purely weak and strong dipolar magnetic field, respectively, as listed in Table~\ref{Table:1}.}
\label{Torque_pure_dipole}
\end{figure*}
\begin{figure*}
\centering
\vspace{0.2cm}
\includegraphics[width=18cm]{\figurepath/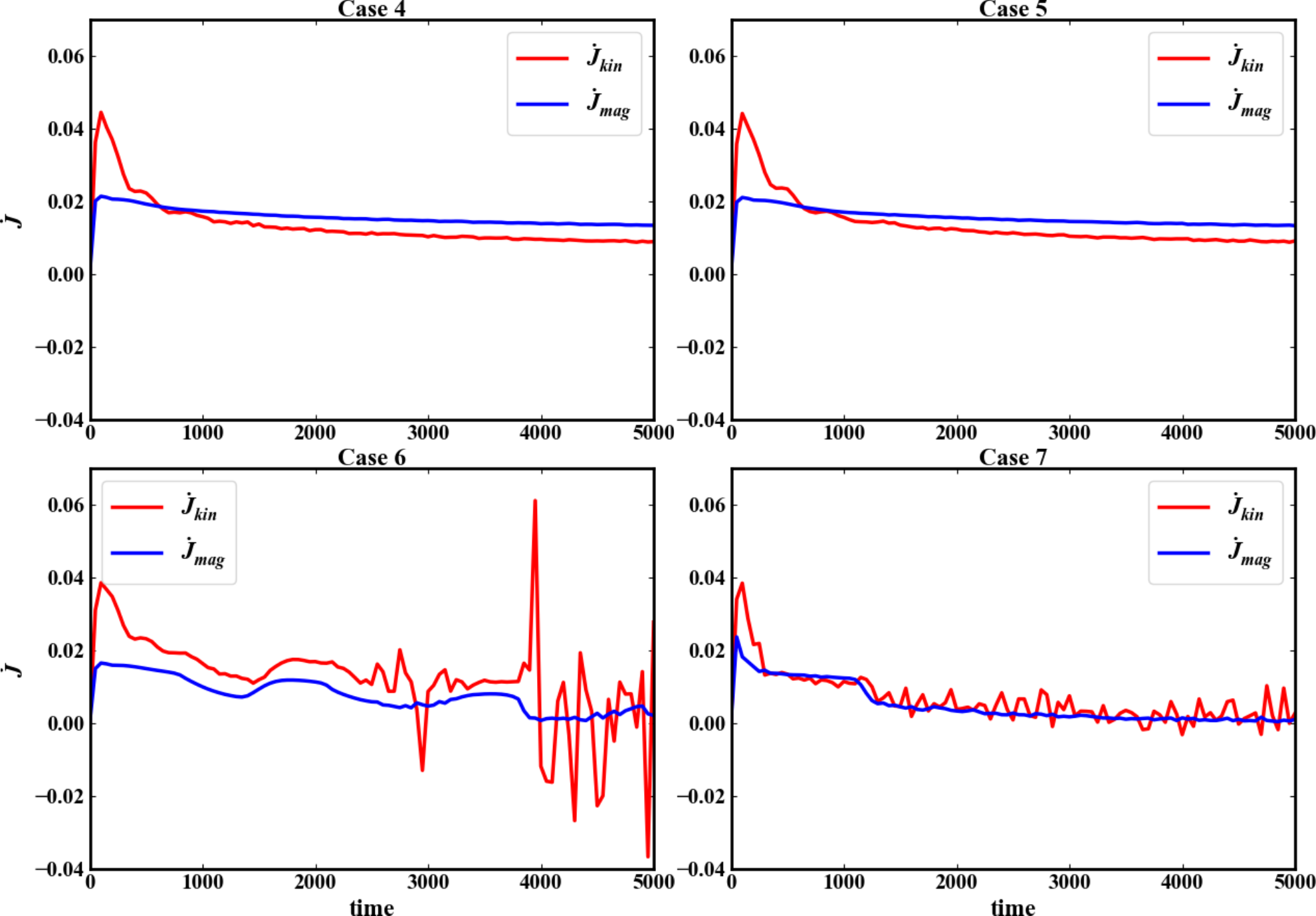}
\caption{Shown are the time evolution of the kinetic $\dot J_{\rm kin}$ and magnetic $\dot J_{\rm mag}$ angular momentum fluxes carried by the outflow, given by Equation \ref{Torque1} and \ref{Torque2}, for runs applying the superposed of the poloidal and dipolar magnetic field, namely, cases 4 to 7, corresponding to dipole field strengths of 1, 10, 100, and 1000, respectively, as listed in Table~\ref{Table:1}.}
\label{Torque_disk_dipole}
\end{figure*}
\section{Accretion-ejection analysis}
\subsection{Mass fluxes}
The most effective way to examine the evolution of both the disk and the launched outflow under various magnetic field configurations is through a quantitative analysis of the accretion and ejection rates across different runs.
Following previous studies of magnetized disk winds ~\citep{2007A&A...469..811Z,2010A&A...512A..82M,2012ApJ...757...65S,2013ApJ...774...12F,2014ApJ...796...29S}, we measure the mass fluxes using a well defined control volume, shown in Figure~\ref{control_volume}.
The control volume extends radially from $1\,r_{\rm i}$ to $10\,r_{\rm i}$ and poloidally from $1.279$ to $\pi/2$ radians, corresponding to approximately three disk scale heights.
Thus, the accretion mass flux is measured by the following integral across the surface ${\rm S}_1$ of the control volume:

\begin{equation}
\dot M_{\rm acc} = \int_{S_1} \rho \, \vec{u}_r \cdot d\vec{s}
\end{equation}

and, the ejection mass flux is measured across the surface ${\rm S}_3$ of the control volume using the integral:

\begin{equation}
\dot M_{\rm ejec} = \int_{S_3} \rho \, \vec{u}_p \cdot d\vec{s}.
\end{equation}

First, we consider the mass fluxes for the reference run, threaded by the initial poloidal magnetic field. Figure~\ref{flux_case1} shows the time evolution of the accretion mass flux ($\dot M_{\rm acc}$) and the outflow mass flux ($\dot M_{\rm ejec}$), evaluated within the control volume defined in Figure~\ref{control_volume}. The solid lines represent the time averaged values, which in this case are similar to the instantaneous values.

From the figure, we observe that the time evolution of the accretion and outflow mass fluxes is smooth, and approximately 26\% of the accreting mass is delivered into the outflow. The evolution of the mass fluxes in the reference run provides further confirmation that the reference run has reached a steady state.
For comparison, we employ the same control volume to measure the mass fluxes for the other runs listed in Table~\ref{Table:1}.
The results are displayed in Figures~\ref{flux_case2_and_case3} and~\ref{flux_case4_to_case7}.

Figure~\ref{flux_case2_and_case3} shows the time evolution of the accretion mass flux ($\dot M_{\rm acc}$) and the outflow mass flux ($\dot M_{\rm ejec}$) for Cases 2 and 3, corresponding to the weak and strong dipolar magnetic fields, respectively, evaluated within the control volume defined in Figure~\ref{control_volume}.
The solid lines indicate the time averaged values of the mass fluxes.

Figure~\ref{flux_case4_to_case7} illustrates the same quantities for Cases 4 through 7, which include the superposition of both the poloidal and stellar dipolar magnetic fields. The dipole strength increases from $1$ to $1000$ across these cases. From Figures~\ref{flux_case2_and_case3} and~\ref{flux_case4_to_case7}, we obtain the following results.

First, including a stellar dipolar magnetic field induces fluctuations in both the disk and the jet, clearly reflected in the evolution of the mass fluxes.
In Cases 2 and 3, with a purely dipolar magnetic field, accretion is suppressed and the subsequent ejection mass fluxes fail to be established, indicating that a weak dipolar field cannot efficiently remove angular momentum or sustain accretion.
Consistently, \citet{2024MNRAS.528.2883Z} found, in 3D ideal MHD simulations of magnetospheric accretion onto a non-rotating star, that a predominantly closed magnetic field topology leads to weak disk outflows.

Increasing the dipolar field strength in Case 3 slightly enhances the accretion mass flux, but the values remain below those of the reference run (see Figure~\ref{flux_case2_and_case3}, left panels).
This indicates that, in addition to the intrinsic inefficiency of the pure dipolar field configuration, the field strength, particularly in Case 2, is too weak to provide the proper torque required to sustain accretion.

Accordingly, Figure~\ref{flux_case4_to_case7} demonstrates that in runs with a smaller contribution from the stellar dipolar magnetic field, such as Case 4 or Case 5, the evolution of the accretion and outflow mass fluxes closely approaches that of the reference run and is smooth.
This indicates that the initial poloidal disk magnetic field efficiently regulates accretion and stabilizes the flow.
In contrast, in Case 7, dominated by the stronger dipolar magnetic field, the accretion rate is reduced and fluctuations are more pronounced.
This reflects the less efficient angular momentum removal and the stronger fluctuations produced by the stellar dipolar magnetic field.

In summary, the evolution across all simulations highlights that the disk magnetic field primarily regulates accretion and stabilizes the mass fluxes, while the stellar dipolar magnetic field mainly introduces fluctuations and variability into the system.
\subsection{Torques and angular momentum fluxes}

One of the key aspects in studying outflow launching from the underlying disk is the analysis of the forces and, consequently, the torques (angular momentum fluxes) involved in this process.
This analysis allows us to assess the effectiveness of various magnetic field configurations in removing angular momentum from the system.

The primary forces acting on the disk-jet system that contribute to its dynamical evolution are gravity, the pressure gradient, and the Lorentz force.
In particular, due to the axisymmetric assumption in our model, angular momentum transport is mainly controlled by the magnetic and kinetic torques, which arise from the Lorentz force and the bulk motion of the flow, respectively.

Consequently, we evaluate the angular momentum fluxes corresponding to the magnetic and kinetic torques for all simulations presented in Table~\ref{Table:1}.

We integrate the torques exerted on the disk by the outflow employing the same control volume defined in  Figure~\ref{control_volume} and the angular momentum fluxes, carried by the outflow are given by,
\begin{equation}
 \dot J_{\rm{kin}} =- \int_{S_3} r\left( \rho v_\phi \vec{v_p}\right)\cdot \vec{ds}
 \label{Torque1}
\end{equation}

\begin{equation}
 \dot J_{\rm{mag}} = \int_{S_3} r\left( \frac{1}{4 \pi} B_\phi \vec{B_p}\right)\cdot \vec{ds}
\label{Torque2}
\end{equation}

where  $\dot J_{\rm kin}$ and $\dot J_{\rm mag}$ are the kinetic and magnetic angular momentum fluxes carried by the outflow, respectively.

Figure~\ref{Torque_ref} displays the time evolution of the kinetic and magnetic angular momentum fluxes carried by the outflow for the reference run.
From the figure, we observe a smooth evolution of both fluxes, confirming that the reference run launches a well collimated, stable outflow from the disk that reaches a steady state, consistent with the results discussed in previous sections.

The kinetic torque is slightly larger during the early evolution and becomes smaller at later times.
This indicates that the bulk motion of the outflow plays a primary role in angular momentum removal initially, while later the magnetic torque becomes more significant.
This suggests that the disk poloidal magnetic field configuration provides a large lever arm and thus an effective magnetic torque.
Consequently, angular momentum is efficiently removed from the disk, supporting accretion.
Consequently, angular momentum is efficiently removed from the disk, supporting accretion. This efficient mass loading along the magnetic field lines enhances the magnetic torque and angular momentum extraction, thereby sustaining the accretion process.

We now consider the evolution of torques and angular momentum fluxes in simulations that include the stellar dipolar magnetic field.
Its presence noticeably changes the overall behavior of the accretion-ejection system.
Figure~\ref{Torque_pure_dipole} displays the time evolution of the torques for the pure dipolar field runs (Cases 2 and 3, weak and strong fields, respectively).
Both the kinetic and magnetic angular momentum fluxes exhibit noticeable fluctuations. In particular, compared to the reference run, the kinetic torque evolution in both cases shows stronger fluctuations, reflecting enhanced variability in the disk dynamics.
The perturbations are even more pronounced in the run with the strong dipolar magnetic field (Case 3).
This behavior is consistent with the fluctuations observed along the magnetic field lines in Figure~\ref{density_Pure_dipole}, as well as with the perturbed disk structure seen in the accretion profile of the strong dipole run (see Figure~\ref{flux_case2_and_case3}).

Additionally, the magnetic and kinetic angular momentum fluxes are substantially lower than in the reference run, indicating that the dipolar magnetic field is inefficient at removing angular momentum and launching the outflow.
Moreover, since a typical MHD disk wind driven jet does not form in these runs, the kinetic angular momentum flux carried by the outflow is also reduced compared to the reference run.
Specifically, the magnetic torque in Case 2 (weak dipolar field) is significantly smaller than in Case 3.

Finally, we consider the simulations in which we implemented a superposition of the initial poloidal and stellar dipolar magnetic fields, namely Cases 4, 5, 6, and 7, shown in Figure~\ref{Torque_disk_dipole}.

The magnetic and kinetic angular momentum fluxes reach higher values than in the runs with a purely dipolar magnetic field. For instance, the time evolution of the angular momentum fluxes in Cases 4 and 5, which include a quite weak dipolar magnetic field, remains smooth and closely resembles that of the reference run.
However, the kinetic angular momentum flux is slightly larger than in the reference run for these two cases (see Figure~\ref{Torque_ref}).

As the dipolar field strength increases, perturbations begin to appear in the kinetic torque, indicating enhanced dynamical variability within the disk. The amplitude of these fluctuations is larger for the runs with the stronger dipolar magnetic field (Cases 6 and 7).
This behavior suggests that stronger dipolar fields induce perturbations that affect angular momentum removal and, consequently, the overall accretion-ejection dynamics.

In conclusion, the presence of a poloidal magnetic field of the disk stabilizes the system and drives the formation of a typical MHD disk wind jet, resulting in smoother and more steady disk-jet evolution.
In addition, we find that the stellar dipolar magnetic field can destabilize angular momentum transport and purely dipolar magnetic field, fails to produce a comparable jet.

\section{conclusions}
We have presented the results of MHD simulations investigating the launching of outflows from a magnetically diffusive, sub-Keplerian accretion disk under the influence of various magnetic field configurations.
Our primary focus was to study how the presence of a dipolar magnetic field impacts the accretion-ejection structure.
The simulations presented are listed in Table~\ref{Table:1} and include Case 1 as the reference model, while Cases 2 and 3 consist of a purely stellar dipolar field, corresponding to weak and strong dipolar magnetic fields, respectively.
Cases 4 to 7 include the superposition of both the stellar dipolar and poloidal magnetic fields, with the dipolar component increasing progressively across these cases.
The simulations were performed in axisymmetry using the MHD code PLUTO 4.4.2 in spherical coordinates.

We have obtained the following results:

\begin{itemize}

\item{First, we presented and discussed the reference simulation, in which the disk is threaded by an initial large scale poloidal magnetic field. The simulation was running for 20000~$t_{\rm i}$ dynamical time, corresponding to ~3100 revolutions at the inner disk radius $r_{\rm i}$.
Both the disk and the jet reached the steady state and a well developed jet consistent with the MHD disk wind model formed.
This confirms the robustness of our model setup and allows us for further using this run as the reference run, to study the impact of implementing a stellar dipolar magnetic field in the setup.
}

\item{
Additionally, we presented simulations threaded exclusively by the  stellar dipolar magnetic field.
Two configurations were considered, corresponding to weaker and stronger dipolar field strengths
Our findings show that although the dipolar magnetic field opens the field lines and allows for low mass loading from the disk, it is  not sufficient for efficient angular momentum removal, resulting in weak accretion.
The outflow also doesn't show a proper acceleration and collimation, consistent with a magnetic torque that is too weak to launch a well structured and collimated jet.
In run with the the strong dipolar field, additional instabilities develop along the field lines near the inner disk, likely triggered by disk perturbations.
These instabilities weaken as the dipole field strength decreases.}

\item{
Moreover, we presented simulations including both a stellar dipolar magnetic field and a poloidal magnetic field anchored in the disk.
In this set of simulations, the magnitude of the stellar dipolar field was gradually increased while the relative contribution of the disk magnetic field was reduced. The results reveal a strong dependence of the disk/jet evolution on the balance between the stellar and disk magnetic fields.
We find that as the stellar dipolar field becomes comparable or stronger than the disk magnetic field, perturbations develop in both the disk and the outflow structure.
Furthermore, accretion is not formed properly and the ejected materials exhibit irregular, distorted structures.
Existence of the initial disk poloidal field alongside the stellar dipole increases both kinetic and magnetic torques relative to the purely dipolar field.
Our results show that in all runs including the dipolar magnetic field the fluctuations in the dynamical evolution of the kinetic and total torque appears  which  is enhanced by increasing the dipolar magnetic field strength.
These results indicate that while the stellar dipolar field can destabilize angular momentum transport, the presence of the disk magnetic field helps stabilizing the system, promoting smoother and more steady accretion-ejection behavior.
Our findings demonstrate that although the magnitude of stellar dipole plays an important role in the evolution of the accretion-ejection structure, the large scale poloidal magnetic field threading the disk is essential for launching and sustaining a stable and well collimated magnetized jet.}

\item{Furthermore, we considered the time evolution of accretion and outflow mass fluxes to study the efficiency of different magnetic field configurations in launching MHD jets.
Our results show that the initial poloidal magnetic field of the disk primarily regulates accretion and stabilizes mass flux evolution, while the stellar dipolar magnetic field modifies the accretion-ejection structure and introduces additional perturbations and variability, with its influence becoming more pronounced as the dipole strength increases.}

\item{
Torque analysis shows that the disk poloidal magnetic field provides a large lever arm for efficient angular momentum removal, enabling steady accretion and a stable jet in the reference run. A purely stellar dipolar field produces strong fluctuations, lower torques, and no significant jet. Superposition runs reveal that even a weak disk field suppresses dipolar instabilities, but strong dipolar fields reintroduce perturbations and reduce ejection efficiency. Thus, the disk poloidal field is essential for stabilizing the system and driving a well collimated outflow.
}
\item{
Our findings are consistent with those of \citet{2000A&A...363..208F}, who showed that strong stellar dipolar magnetic fields can trigger instabilities along the rotation axis and are unlikely to produce stable jets. Our results are also in agreement with \citet{2009ApJ...692..346F}, who studied configurations combining a disk poloidal field with a stellar dipolar field and found that strong stellar winds lead to weakly collimated outflows. Together, these studies support the conclusion that well collimated protostellar jets are primarily driven by disk winds rather than by the stellar magnetosphere alone.
}

\item{Furthermore, we performed additional test simulations that explicitly include an inner gap representing the disk truncation region (see Appendix).
These simulations were designed as a robustness test of the treatment of the innermost disk region.
The results are fully consistent with those of our main simulations, confirming that our main conclusions are not sensitive to the explicit treatment of the inner truncation region.
The magnitude of the stellar dipolar magnetic field nevertheless remains an important parameter, affecting the magnetospheric structure and the associated outflow.
Within the present setup, however, the stellar dipole alone does not produce the persistent and well collimated outflow obtained when the disk is threaded by a large scale poloidal magnetic field.
A fully self-consistent investigation of a rotating stellar magnetosphere, including stellar rotation and an appropriate conducting stellar boundary, is beyond the scope of the present study and is left for future work.}
\end{itemize}

\begin{acknowledgements}
We thank Andrea Mignone and the PLUTO team for the possibility of using their code. We would also like to acknowledge the helpful and insightful comments from an anonymous referee, which have greatly contributed to productive discussions and an improved presentation of our findings. We would also thank Christian Fendt for the helpful
comments.  Our simulations were performed on the TURIN cluster of the Institute for Research in Fundamental Sciences (IPM) and the Gavazang cluster of the Institute for Advanced Studies in Basic Sciences (IASBS) as well as the VERA cluster of Max Planck Institute for astronomy(MPIA).

\end{acknowledgements}

\appendix

\section{Test runs including the inner gap}
\label{inner gap runs}

We performed additional test simulations that explicitly include an inner gap representing the disk truncation region. These simulations are listed in Table~\ref{Table:2}.
Here, we present the test runs that include the inner gap between the disk and the central star. Similar to our main simulations, the inner boundary conditions described in Section~\ref{section:BC} are applied at $R=1$ as an ``internal boundary'' defined by the PLUTO code.
In the gap region between the inner disk radius and the central object, the domain is initially filled with a corona density and pressure distribution, consistent with the magnetospheric truncation scenario.
This setup ensures a meaningful comparison with our main results.

We emphasize that these simulations are intended as a robustness test of the treatment of the innermost disk region, rather than as a fully self-consistent model of the stellar magnetosphere.
A complete treatment of the disk-magnetosphere interaction would additionally require stellar rotation and an appropriate conducting boundary condition at the stellar surface, allowing the magnetic coupling, field twisting, and angular momentum exchange between the star and the disk to be treated self consistently.
The purpose of the present test runs is instead to examine whether our main accretion-ejection structure and conclusions depend sensitively on explicitly resolving a finite disk truncation region.

\begin{figure}
\centering
\includegraphics[width=18cm]{\figurepath/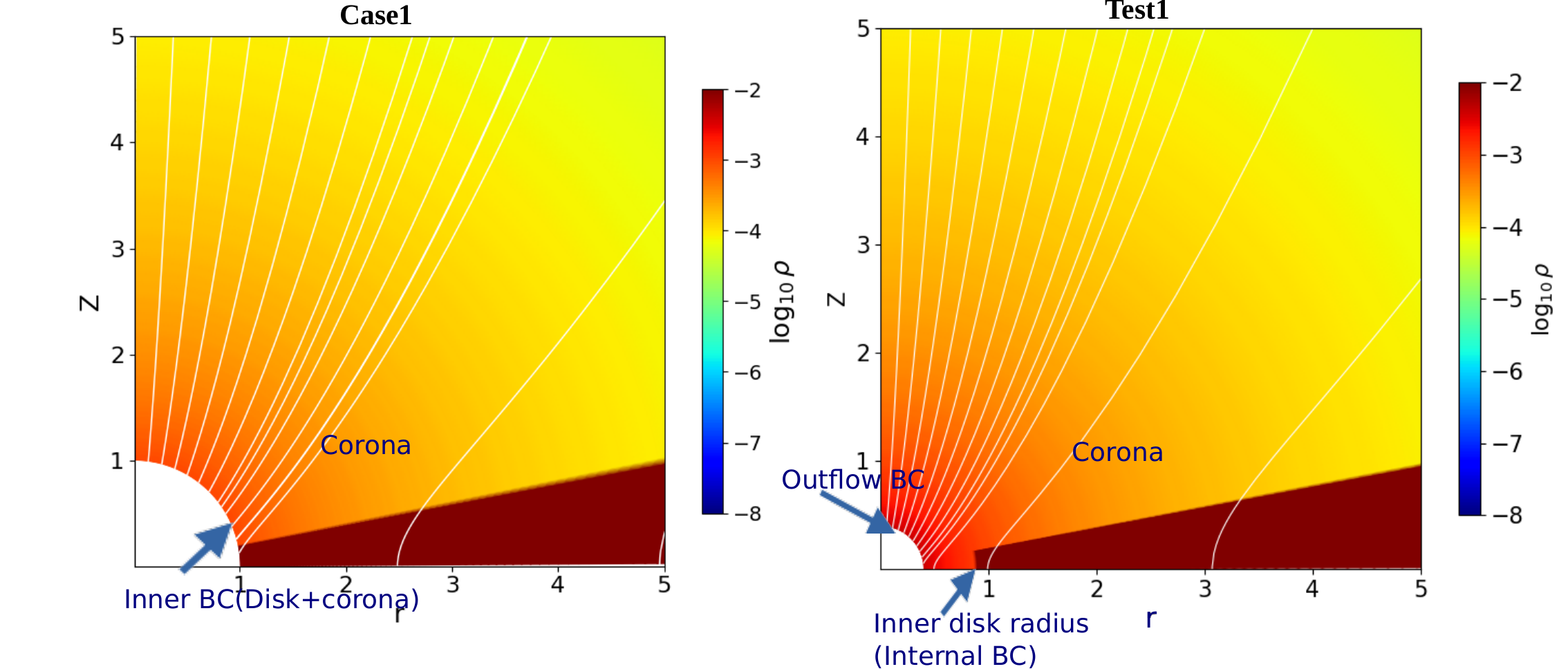}
\caption{The inner regions of the main reference run (Case1, left panel) and the new test run (Test1, right panel), including the initial gap between the disk and the central star, are shown at the initial time.
In the main reference run (case1, left panel), the inner disk edge coincides with the inner radial boundary of the computational domain at $R=1$, and the boundary conditions for both the disk and corona are directly applied at this location.
In contrast, in the test run (Test1, right panel), the inner radial boundary of the computational domain is located at $R=0.4$ and is treated with a standard outflow boundary condition, where the variables are copied into the ghost cells. The inner edge of the disk is defined as an internal boundary at $R=1$.
The region between the central star and the inner disk edge is initially filled with the prescribed coronal density and pressure distribution.}
\vspace{0.2cm}
\label{rho_gap_zoom_comparision}
\end{figure}

For comparison, Figure~\ref{rho_gap_zoom_comparision} shows the innermost regions of the main reference run (Case1, left panel) and the new test run (Test1, right panel) at the initial time, including the gap between the disk and the central star.
In these test models, the computational domain extends inward to $R=0.4$, while the accretion disk begins at $R=1$, thus explicitly resolving the inner gap between the central object and the disk.
Considering the figure, in run ``Test1'', the inner radial boundary of the computational domain is located at $R=0.4$ and is treated with a standard outflow boundary condition, where the variables are copied into the ghost cells.
The inner edge of the disk at $R=1$ is treated as an internal boundary, where the disk boundary conditions are applied.
The region between the central star and the inner disk edge is initially filled with the prescribed coronal density and pressure distribution.
In contrast, in the main reference run (Case1), the inner disk edge coincides with the inner radial boundary of the computational domain at $R=1$, and the boundary conditions for both the disk and corona are directly applied at this location without an explicitly resolved inner gap.

When the disk is initially threaded by a large scale poloidal magnetic field, a stable, well collimated jet is launched and sustained, demonstrating that the inclusion of the inner gap does not alter the disk-wind launching mechanism. In contrast, when only a stellar dipolar magnetic field is present, no sustained MHD disk wind or well collimated jet is produced.
Finally, when both magnetic field components are included, the outcome depends on their relative strengths.
A dominant stellar dipolar field suppresses the formation of a large scale MHD disk wind, whereas reducing the relative strength of the dipolar component allows the disk threading poloidal field to drive a stable, well collimated jet.

The strength of the stellar dipolar magnetic field nevertheless remains an important parameter in determining the magnetospheric structure and the associated outflow. In the present setup, increasing the dipole strength modifies the accretion-ejection structure and can produce stronger magnetospheric effects; however, the stellar dipole alone does not produce the persistent and well collimated outflow obtained when the disk is threaded by a large scale poloidal magnetic field.
Thus, our results distinguish between the influence of the stellar dipole on the magnetospheric flow and the ability of the large scale magnetic field threading the disk to sustain a persistent, well collimated MHD outflow.

A fully self consistent treatment of the stellar magnetosphere, including stellar rotation and an appropriate conducting stellar boundary, could introduce additional magnetic twisting, angular-momentum exchange, and time-dependent magnetic field evolution in the inner gap.
Such effects may modify the outflow structure close to the star and are not captured by the present gap tests. They do not, however, alter the purpose of these simulations, which is to test the robustness of the large scale accretion-ejection structure against the explicit treatment of the inner truncation region.

These results are fully consistent with those presented in Section~3 of the manuscript.
We therefore conclude that explicitly resolving the magnetospheric truncation region does not qualitatively alter the main conclusions of this work. In particular, our principal result remains unchanged: the long term launching and collimation of the jet are governed primarily by the large scale poloidal magnetic field threading the disk, while a stellar dipolar magnetic field alone is insufficient to sustain a persistent, well collimated large scale MHD jet.

\begin{figure}
\centering
\includegraphics[width=18cm]{\figurepath/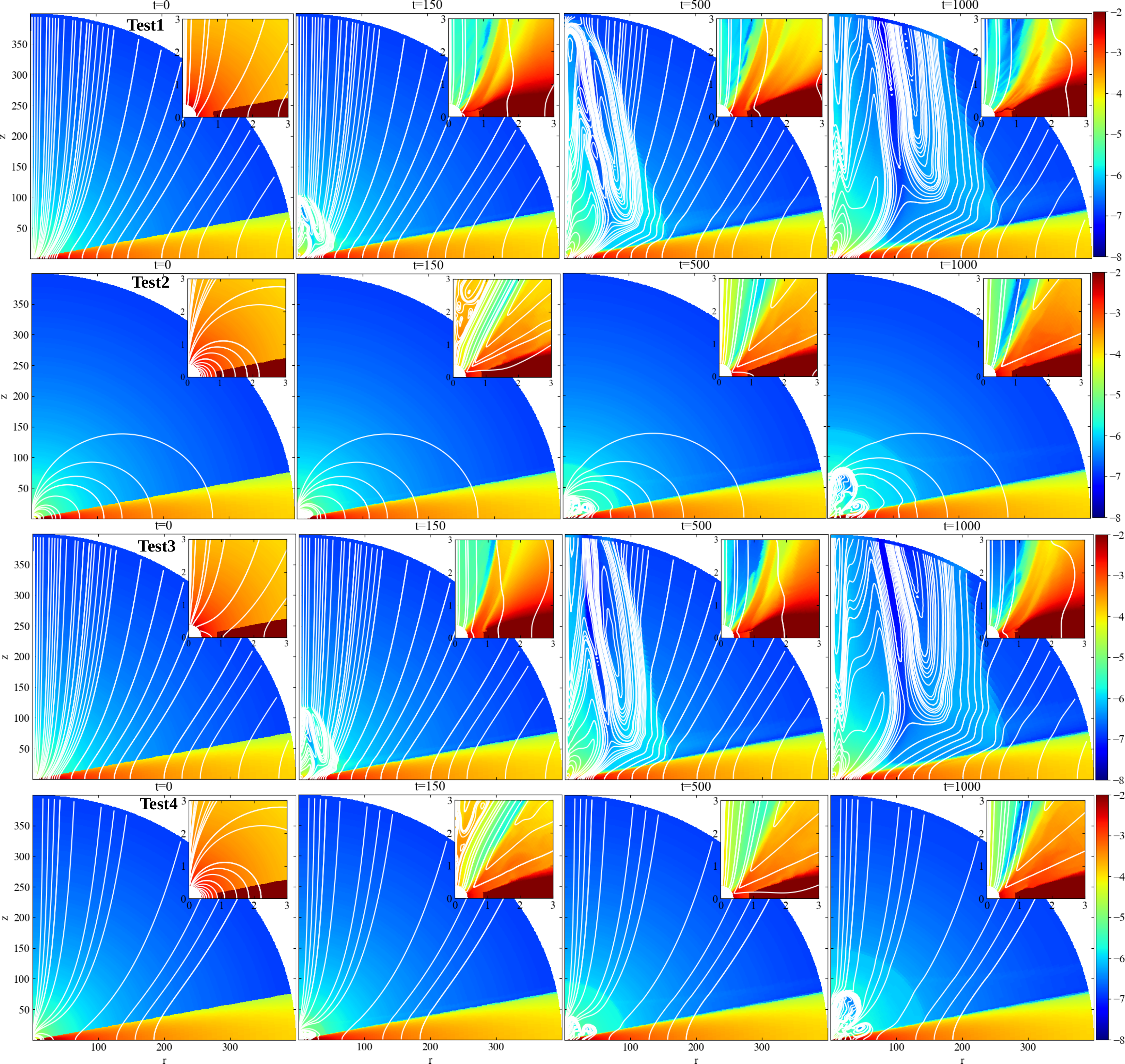}
\caption{Snapshots of the mass density distributions for all test runs including an inner gap between the disk and the central star, as listed in Table~\ref{Table:2}. The full computational domain is shown here. The solid lines represent the magnetic field lines.}
\vspace{0.2cm}
\label{rho_gap_all}
\end{figure}

\begin{figure}
\centering
\includegraphics[width=14cm]{\figurepath/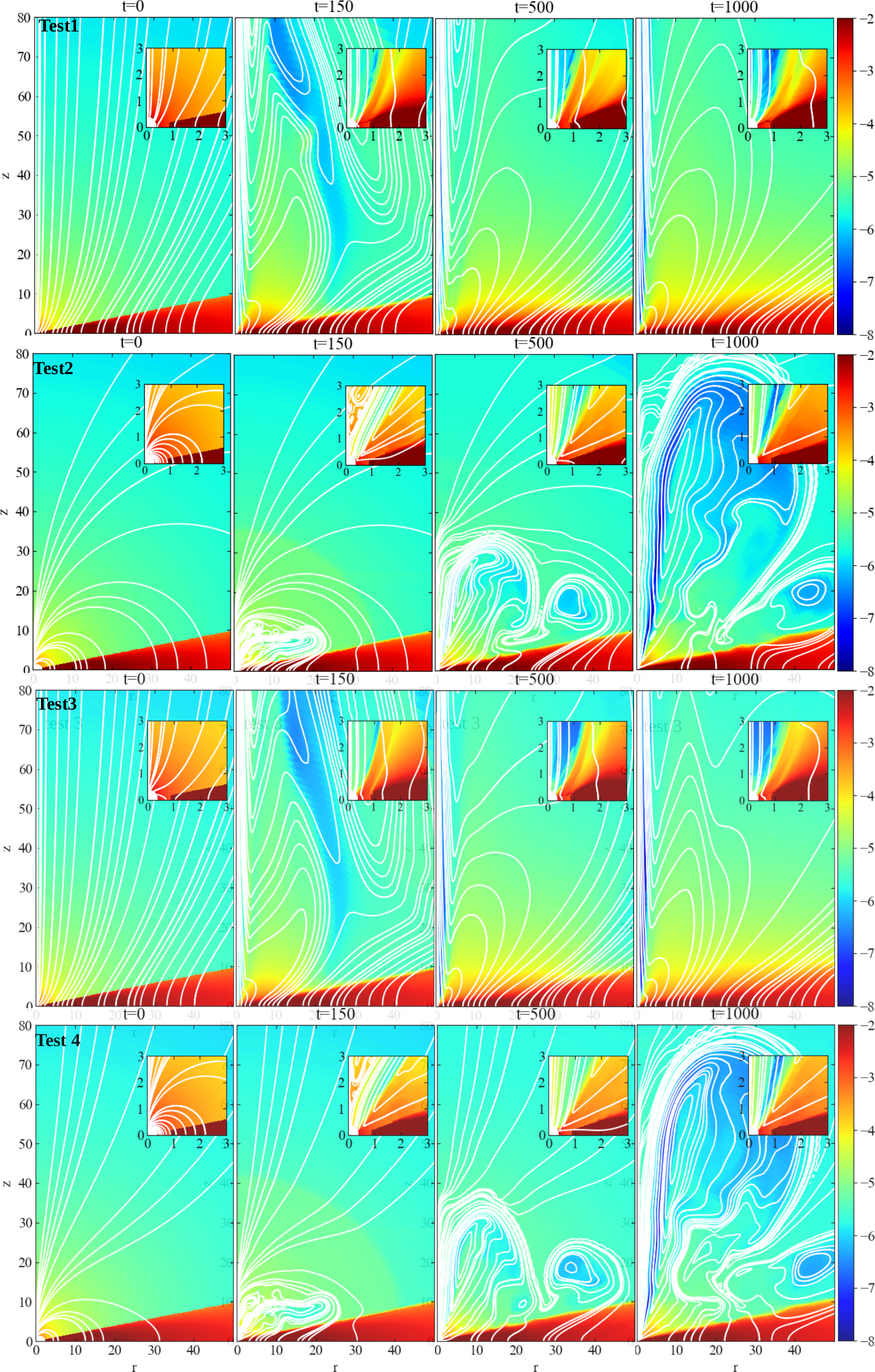}
\caption{Snapshots of the mass density for the test runs including an inner gap between the disk and the central star, as listed in Table~\ref{Table:2}. The displayed region corresponds to the inner part of the system.}
\vspace{0.2cm}
\label{rho_gap_zoom_all}
\end{figure}

\begin{figure}
\centering
\includegraphics[width=17cm]{\figurepath/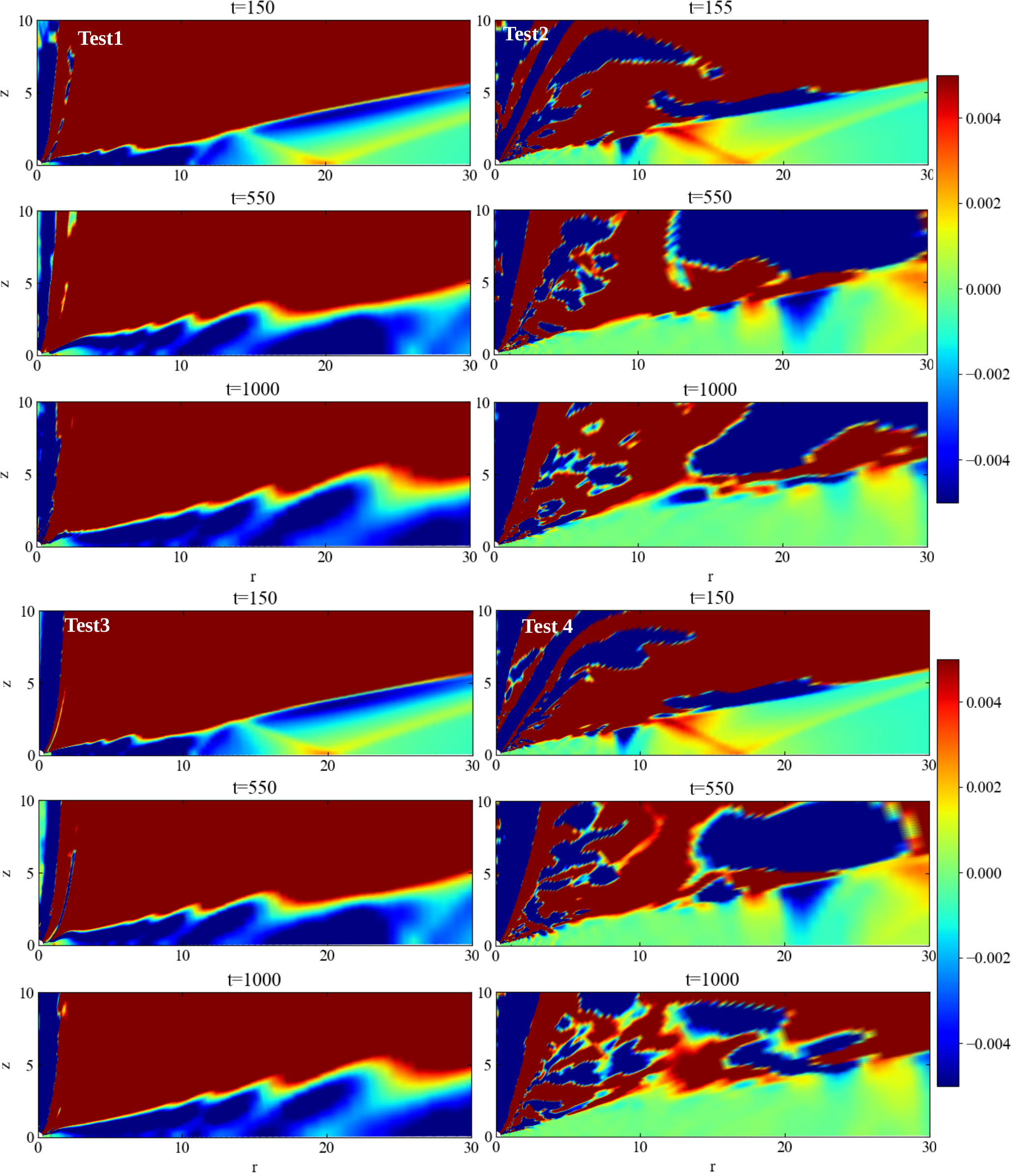}
\caption{Snapshots of the radial velocity inside the disk are shown for the test runs that include the inner gap between the disk and the central star as listed in Table \ref{Table:2}.
The displayed region corresponds to the innermost part of the system.
}
\label{vr_gap_all}
\end{figure}

Figures~\ref{rho_gap_all} and \ref{rho_gap_zoom_all}, display the full domain and the inner region of the systems, for test runs as listed in Table \ref{Table:2}, respectively.
Additionally, Figure \ref{vr_gap_all} illustrates the radial velocity within the disk for the same set of runs.
Considering these figures, we find that the results are in excellent qualitative agreement with our main simulations(presented in section~\ref{sec:results_main_runs}).

\begin{table}
\centering
\caption{Characteristic parameters of the simulation runs. The listed test runs account for the inner gap between the disk and the central star. In Test1 a purely poloidal disk magnetic field is used, while Test2 employs a purely dipolar field. Test3 incorporates both dipolar and poloidal components, and Test4 includes a dipolar field along with a weaker poloidal component.}
\begin{tabular}{|c|c|c|c|c|c}
\hline
\hline
RunID & B Disk & B Dipole & Vector potential  \\
\hline
\hline
Test1 &Yes & No &	$A_{\phi,d}$ \\
\hline
Test2 & No &	Yes &	$100A_{\phi,dp}$  \\
\hline
Test3 & No & Yes &	$100A_{\phi,dp}+A_{\phi,d}$ \\
\hline
Test4  &Yes &Yes &$100 A_{\phi,dp}+0.001A_{\phi,d}$ \\
\hline
\hline
\end{tabular}
\label{Table:2}
\end{table}


\section{Test of the inner outflow boundary condition}
\begin{figure*}
\centering
\includegraphics[width=18cm]{\figurepath/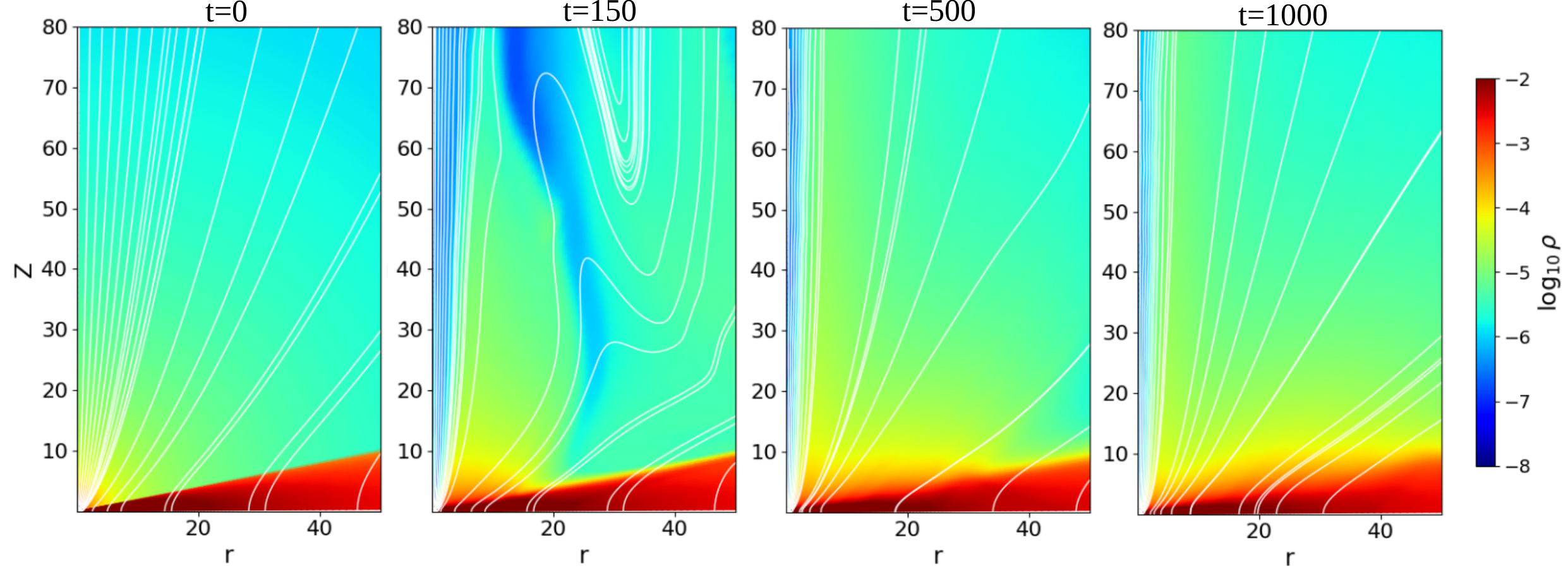}
\caption{Shown are the snapshots of the mass density for the test run employing a standard outflow boundary condition at the inner radial boundary, $R_{\rm in}$, at dynamical times $t=0$, $150$, $500$, and $1000$. The displayed region corresponds to the inner part of the system.}
\vspace*{0.5 cm}
\label{rho_outflow_BC_inner}
\includegraphics[width=17cm]{\figurepath/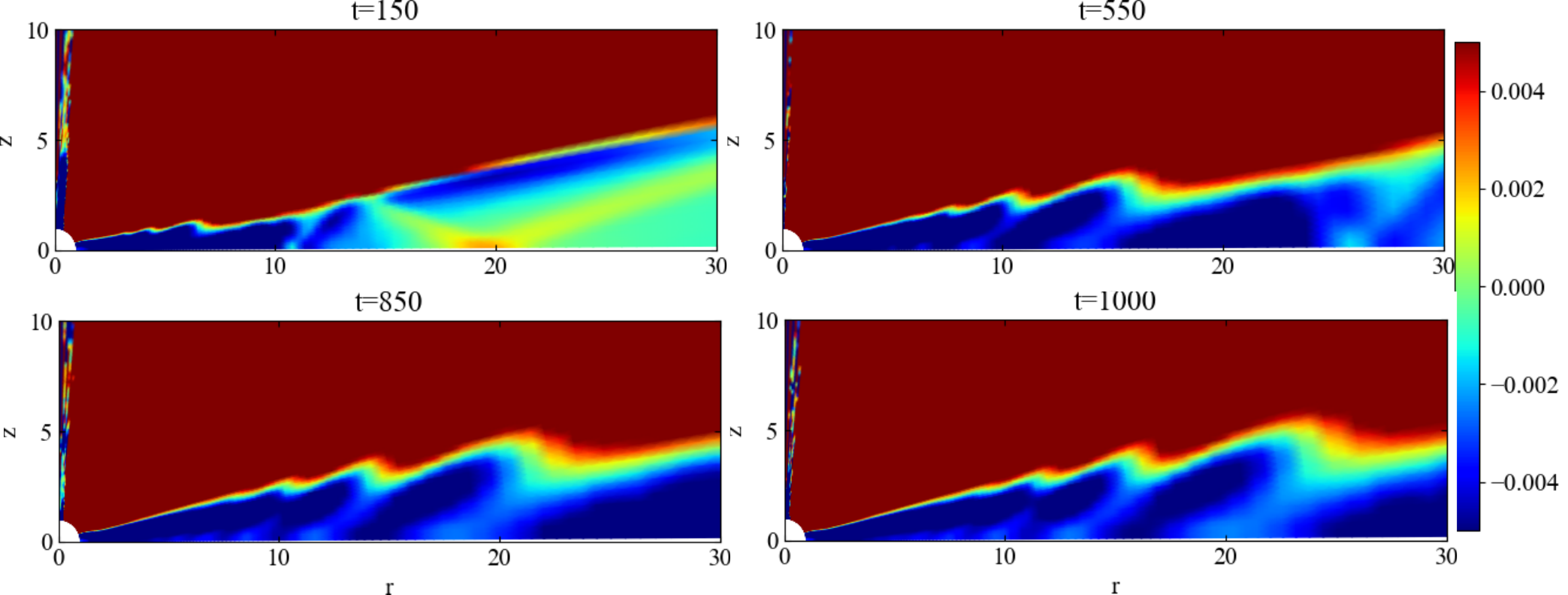}
\caption{Shown are snapshots of the radial velocity for the test run employing the standard outflow boundary condition at the inner radial boundary, $r_{\rm in}$. The displayed region corresponds to the inner part of the system.
}
\vspace*{0.5 cm}
\label{vr_outflow_BC_inner}
\end{figure*}

To assess the sensitivity of our results to the boundary treatment, we performed an additional simulation using a standard outflow boundary condition at the inner radial boundary of the computational domain, while applying the poloidal magnetic field threading the disk.
Snapshots of the radial velocity and mass density, from this test are shown in Figures~\ref{vr_outflow_BC_inner} and \ref{rho_outflow_BC_inner}, respectively.

The comparison shows that the overall evolution remains qualitatively unchanged.
In particular, the accretion disk develops in a similar manner, and a magnetically driven jet is launched smoothly with comparable morphology.
This confirms that the jet launching mechanism is governed primarily by the disk dynamics and the magnetic field configuration rather than by the specific implementation of the inner boundary condition.

The main difference is that the standard outflow boundary permits a continuous loss of mass through the inner boundary, resulting in a gradual depletion of the computational domain and a reduction of the numerical time step during the simulation.
Apart from these expected numerical effects, no significant differences are found in the large scale disk or jet structure.

\section{additional plots}

\begin{figure*}
\centering
\includegraphics[width=17cm]{\figurepath/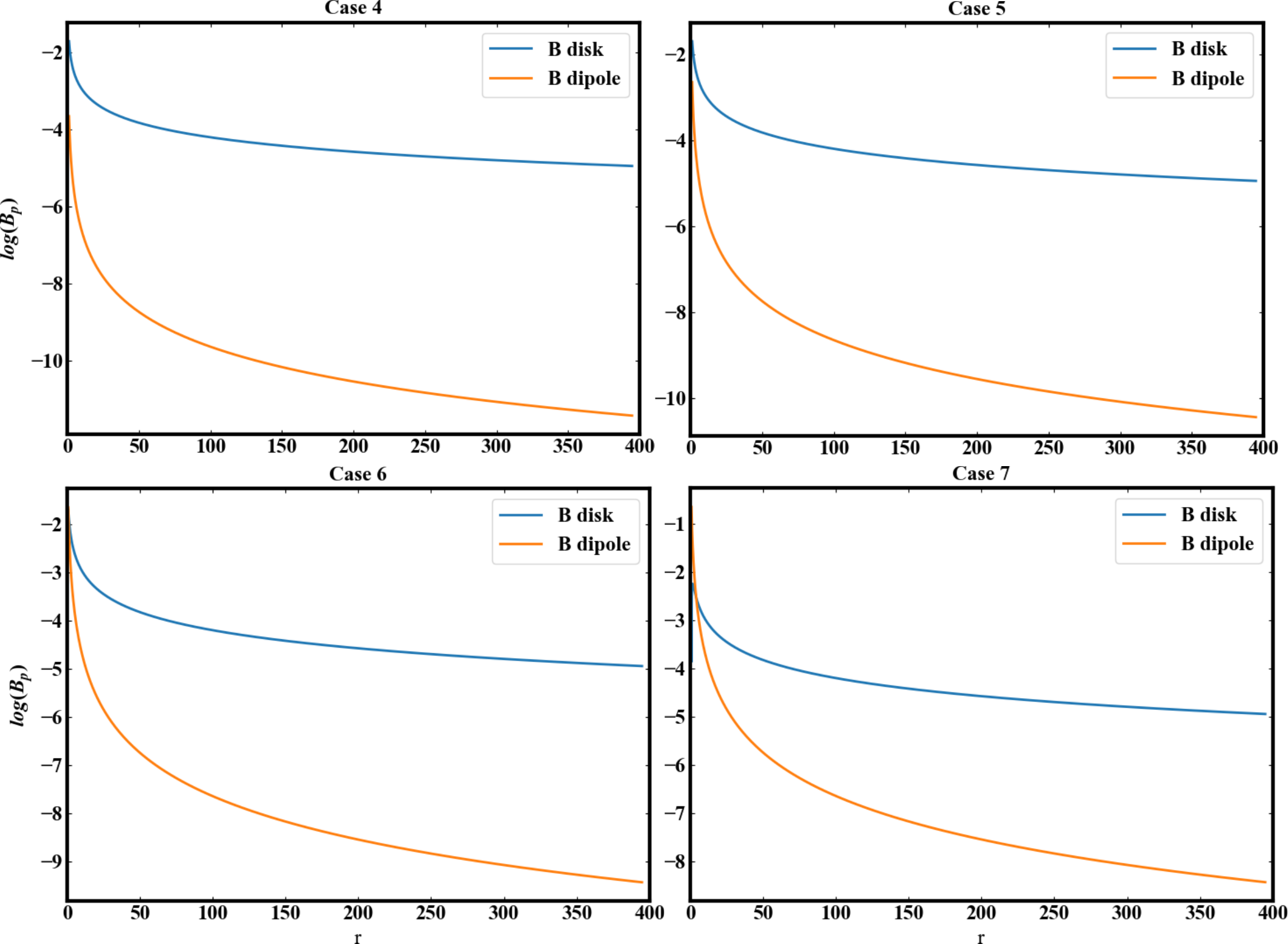}
\caption{Shown are the radial profiles of the initial magnetic field strength on a logarithmic scale ($\log B_{\rm p}$) for the dipolar stellar magnetic field and the poloidal disk magnetic field in cases 4 to 7, as listed in Table~\ref{Table:1}}
\label{Bp_strength_comp1}
\end{figure*}

\bibliographystyle{aasjournal}
\bibliography{myref_Feb3_2026}

\end{document}